\documentclass[12pt,a4paper]{article}

\usepackage{jheppub}
\usepackage[usenames,dvipsnames]{xcolor}

\usepackage{amssymb} 
\usepackage{amsmath}
\usepackage{mathtools}
\usepackage{amsfonts}    
\usepackage{dsfont}
\usepackage{pdfpages}
\usepackage{verbatim}
\usepackage{stmaryrd}
\usepackage{graphicx}

\usepackage{tensor}
\usepackage{mathrsfs}
\usepackage{textgreek} 
\usepackage[mathscr]{euscript}
\usepackage[smalltableaux]{ytableau}
\ytableausetup{centertableaux}
\usepackage{tikz}
\usetikzlibrary{decorations.pathreplacing, decorations.markings,calc,shapes.misc,decorations.pathmorphing,patterns.meta, math}
\usepackage{slashed}
\usepackage{datetime}
\usepackage{hyperref}
\hypersetup{
	pdfencoding=unicode,
	colorlinks=true,
	urlcolor=Maroon,
	linkcolor=RoyalBlue,
	citecolor=Maroon,
	pdftitle={The canonical differential equations of the one-loop-like integrals},
	pdfauthor={Jiaqi Chen, Bo Feng, Liang Zhang},
	pdfdisplaydoctitle=true,
	pdfstartview=FitH,
	linktocpage=true
}

\usetikzlibrary{shapes}

\newcommand{\ba}{\begin{align}}
	\newcommand{\ea}{\end{align}}
\newcommand{\be}{\begin{equation}}
	\newcommand{\ee}{\end{equation}}
\def\bd{\begin{tikzpicture}}
	\def\ed{\end{tikzpicture}}

\allowdisplaybreaks

\def\XXint#1#2#3{{\setbox0=\hbox{$#1{#2#3}{\int}$}
		\vcenter{\hbox{$#2#3$}}\kern-.5\wd0}}

\definecolor{light-gray}{gray}{0.75}

\renewcommand\d{\text{d}}

\renewcommand{\leq}{\leqslant}
\renewcommand{\geq}{\geqslant}

\def\O #1{\overline{#1}}
\def\bm #1{\boldsymbol{#1}}

\newcommand{\bea}{\begin{eqnarray}}
	\newcommand{\eea}{\end{eqnarray}}
\newcommand{\bean}{\begin{eqnarray*}}
	\newcommand{\eean}{\end{eqnarray*}}
\newcommand{\nnn}{\nonumber\\}

\title{The canonical differential equations of the one-loop-like integrals}

\author[a,b,c]{Jiaqi Chen,}\emailAdd{jiaqichen@cup.edu.cn}
\author[c,d]{Bo Feng,}\emailAdd{fengbo@csrc.ac.cn}
\author[c]{Liang Zhang\footnote{The correspondence author}}\emailAdd{liangzhang@csrc.ac.cn}

\affiliation[a]{Beijing Key Laboratory of Optical Detection Technology for Oil and Gas, China University of Petroleum-Beijing, Beijing 102249, China}
\affiliation[b]{Basic Research Center for Energy Interdisciplinary, College of Science, China University of Petroleum-Beijing, Beijing 102249, China}
\affiliation[c]{Beijing Computational Science Research Center, Beijing 100084, China}
\affiliation[d]{Peng Huanwu Center for Fundamental Theory, Hefei, Anhui, 230026, China}
\abstract{
	Recently, a new approach for high loop integrals has been proposed  in \cite{Huang:2024nij}, where the whole parameter integration has been divided into two parts: a one-loop-like integration and the remaining parameter integration. In this paper, we systematically study the one-loop-like integrals. We establish the IBP relations for the integral family and show how to complete the reduction. We find the canonical master integrals and write down the corresponding canonical differential equations.    
} 

\begin{document}
	
	\setcounter{tocdepth}{3}
	\maketitle
	\setcounter{page}{2}
	
	\section{\label{sec:intro}Introduction}

	The increasing precision of high-energy physics experiments necessitates correspondingly accurate theoretical predictions. Perturbative quantum field theory provides the primary framework for achieving these high-precision predictions in particle physics. A fundamental, yet challenging, aspect of perturbative calculations lies in evaluating Feynman integrals arising from scattering amplitudes. To address this, numerous systematic methods for computing Feynman integrals have been developed.
	
	Among these methods, Integration-By-Parts (IBP) identities \cite{Chetyrkin:1981qh, Tkachov:1981wb, Laporta:2000dsw} and the differential equation method \cite{Kotikov:1990kg, Kotikov:1991pm, Bern:1993kr, Gehrmann:1999as} derived from them have been proven particularly effective. IBP relations establish linear dependencies among Feynman integrals, enabling the reduction of a large number of integrals to a linear combination of a finite set, known as master integrals. These master integrals, in turn, satisfy a system of first-order linear differential equations, forming the basis of the differential equation method.
	
	For analytic computations, the canonical differential equation (CDE) approach \cite{Henn:2013pwa, Henn:2014qga}, building upon IBP and the differential equation method, represents the most powerful technique currently available. By judiciously selecting master integrals, the differential equations can be transformed into a d$\log$-form proportional to $\epsilon$. When this transformation is fully rationalizable, it facilitates an iterative solution for the analytic expressions of the master integrals at each order in $\epsilon$, typically expressed in terms of multiple polylogarithm (MPL) functions. In cases where full rationalization is unattainable, alternative approaches for obtaining analytic results have been recently developed \cite{He:2023qld,Papathanasiou:2025stn}. Consequently, obtaining the canonical form of the differential equations often signifies, or at least significantly facilitates, the derivation of analytic solutions for the master integrals. However, extending this method beyond MPL functions—for instance, to cases involving elliptic integrals—remains an active area of research.
	
	For numerical computations, the generalized power series expansion method \cite{Moriello:2019yhu, Hidding:2020ytt}, based on IBP and the differential equation approach, offers a systematic and efficient framework for numerically solving differential equations, with convenient packages such as DiffEXP \cite{Hidding:2020ytt} and SeaSyde\cite{Armadillo:2022ugh,Armadillo:2025mvu}. This method,  when combined with d$\log$-form differential equations, has also recently demonstrated success in deriving analytic solutions for tree-level cosmological correlators in curved spacetime \cite{Chen:2024glu}. Solving differential equations also need the boundary conditions as import. For the determination of boundary conditions, one can choose Monte Carlo-based sector decomposition \cite{Hepp:1966eg, Heinrich:2008si}, or numerical differential equations-based methods including the AMFlow method \cite{Liu:2017jxz} and its associated package \cite{Liu:2022chg}, as well as an alternative approach developed in \cite{Chen:2023hmk} and implemented in the AmpRed package \cite{Chen:2024xwt}. Furthermore, alternative methods exist for the computation of Feynman integrals, as demonstrated in one-loop calculations \cite{Phan:2018cnz,Riemann:2019gqf} and extended to multi-loop scenarios \cite{Hidding:2022ycg,Dubovyk:2022frj}.
	
	At first glance, numerical methods based on differential equations appear to offer a systematic approach for computing Feynman integrals. However, for cutting-edge physical processes demanding extremely high precision, these computations are becoming increasingly complex and resource-intensive. In many cases, the IBP reduction process for higher loops becomes the primary bottleneck due to its excessive computational demands, rendering it impractical for these applications. Therefore, the development of novel and efficient methods for computing Feynman integrals remains an urgent and crucial challenge.
	
	Recently, a new approach proposed in \cite{Huang:2024nij,Huang:2024qan} has demonstrated significant potential for accelerating the computation of arbitrary two-loop and higher loop diagrams. In \cite{Huang:2024nij}, leveraging either Feynman parameterization or the Lee-Pomeransky (LP) representation, the authors introduced a novel parameterized representation of Feynman integrals, which we refer to as the Huang-Huang-Ma (HHM) representation. Within this framework, integrals at any loop order can be reformulated into a structure comprising a one-loop-like integral kernel (it is called { fixed-branch integrals} in \cite{Huang:2024nij,Huang:2024qan} and the precise nature of this analogy will be discussed subsequently) and a series of additional integrals.
	
	It has long been recognized that one-loop integrals exhibit significantly greater simplicity compared to their higher-loop counterparts. This inherent simplicity has been extensively studied in the context of symbol alphabets, canonical master integrals, and canonical differential equations for arbitrary one-loop integrals, employing various approaches \cite{Chen:2022fyw, ArkaniHamed:2010gh, Gehrmann:2011xn, Drummond:2013nda, Henn:2013pwa, Arkani-Hamed:2014via, Bern:2014kca, Herrmann:2019upk, Bourjaily:2019exo, Chen:2020uyk, Chen:2022lzr, Feng:2022hyg, Hu:2023mgc}.
	
	In the HHM representation at two loops, beyond the one-loop-like integral kernel, only two additional integrals remain. This observation suggests that if the one-loop-like integrals can be efficiently computed or directly expressed analytically, the entire two-loop integral problem can be reduced to a double integral. In \cite{Huang:2024nij}, the authors implemented a computational scheme based on this concept, demonstrating remarkable efficiency. They introduced an exceptionally effective reduction scheme for the one-loop-like integrals and employed a relatively straightforward numerical differential equation method.
	
	Motivated by this progress, in this paper we will investigate the properties of this one-loop-like integrals  further: not only to enhance the computational efficiency of this approach but also to advance the understanding of the mathematical structures underlying higher-loop integrals. Consistent with the findings of \cite{Chen:2022fyw}, we observe that the one-loop-like integrals  in the HHM representation exhibit properties under IBP and differential equations that closely resemble those of genuine one-loop integrals. Specifically, all of its canonical master integrals, as demonstrated in \cite{Chen:2020uyk, Chen:2022fyw}, can be expressed using two remarkably simple formulas, applicable separately to cases with an odd or even number of propagators.
	
	Moreover, similar to the results in \cite{Chen:2022fyw}, we identify a comparable canonical differential equation structure and symbol alphabet. In non-degenerate cases, we find that a given sector contains only a single master integral and that the corresponding canonical differential equations depend on subsectors with at most two propagators less. This suggests that, in the future, one might attempt to directly derive analytic expressions for arbitrary one-loop-like integrals or employ them for efficient numerical computations.
	
	Although the canonical differential equations we present involve square roots, a simple diagonal transformation can be performed, if necessary, to obtain rational differential equations with respect to any chosen variable. This rational form enhances the efficiency of direct numerical differential equation methods. Additionally, we introduce an alternative reduction scheme for the one-loop-like  integrals, which is expected to yield results comparable to those presented in \cite{Huang:2024nij}, albeit with slight variations.

	The arrangement of the paper is as follows. In section 2, we recall the Feynman parameterization of general loop integrals and how the one-loop-like integrals appear according to the proposal made in \cite{Huang:2024nij}. In section 3, we study the IBP reduction of these one-loop-like integrals. In section 4, we construct the canonical master integrals and their corresponding canonical differential equations. In section 5, some examples of degenerate bases have been discussed. Finally, a conclusion is given in section 6. A further technical point is clarified in appendix A. Appendix B provides a validation of our method at the one-loop level.

	\section{\label{sec:fey}The Feynman parameterization}
	
	In this section, we review the Feynman parameterization form of general loop integrals. The purpose is to set up the framework of one-loop-like integrals of our focus on this paper. We will follow the line presented in \cite{Huang:2024nij}.

	The $L$-loop Feynman integral (FI) is
	\begin{align}
		I  = \int \prod\limits_{r=1}^{L} \frac{d^D l_r}{i \pi^{\frac{D}{2}}} 
		\prod\limits_{j=1}^{n} \frac{1}{D_j^{\nu_j}},
	\end{align}
	where $D$ denotes the spacetime dimension, $D_j$ represents the propagators,  $n$ is the total number of propagators, and $\nu_j$ are positive exponents. Using Feynman parameterization to combine the denominators, we have 
	\begin{align}
		\frac{1}{\prod_{j=1}^n D_j^{\nu_j}}=\frac{\Gamma(\nu)}{\prod_{j=1}^{n}\Gamma(\nu_j)}\int d^n \boldsymbol{a}~\delta(1-\sum_{j=1}^n a_j)\frac{\prod_j a_j^{\nu_j-1}}{(\sum_{j=1}^n a_j D_j)^{\nu}}\,,
	\end{align}
	where $\boldsymbol{a}\equiv(a_1,\ldots,a_n)$ is the list of the Feynman parameters, and $\nu= \nu_1+\ldots+\nu_n$. Since the inverse propagators are $D_j=-q_j^2+m_j^2$, where $q_j$ is a linear combination of loop and external momenta, the denominator can then be expressed as
	\begin{align}
		\sum_{j=1}^n a_j D_j=-\sum_{r=1}^L\sum_{s=1}^L l_r \boldsymbol{K}_{rs} l_s+\sum_{r=1}^L 2l_r\boldsymbol{v}_r+J\,,~~~\label{2.3}
	\end{align}
	where $\boldsymbol{K}$ is an $L\times L$ matrix, $\boldsymbol{v}$ is an $L\times 1$ column matrix, and $J$ is a scalar. Using \eqref{2.3}, the Symanzik polynomials $\mathcal{U}$ and $\mathcal{F}$ define as
	\begin{align}
		\mathcal{U}=\text{det}(K)\equiv |\boldsymbol{K}|, \quad\quad \mathcal{F}=|\boldsymbol{K}|(J+\boldsymbol{v}^T \boldsymbol{K}^{-1} \boldsymbol{v})\,,
	\end{align}
	After integrating over the loop momentum, the FI can be written as
	\begin{align}\label{eq:FeyPara}
		I=\frac{\Gamma(\nu-\frac{LD}{2})}{\prod_{j=1}^n\Gamma(\nu_j)}\int d^n \boldsymbol{a}~\delta(1-\sum_{j=1}^n a_j)(\prod_{j=1}^n a_j^{\nu_j-1})\frac{[\mathcal{U}(\boldsymbol{a})]^{\nu-\frac{(L+1)D}{2}}}{[\mathcal{F}(\boldsymbol{a})]^{\nu-\frac{LD}{2}}}\,
	\end{align}
	For general integrals with numerator ${\cal N}(l)$
	\begin{align}
		I  = \int \prod\limits_{r=1}^{L} \frac{d^D l_r}{i \pi^{\frac{D}{2}}} 
		\prod\limits_{j=1}^{n} \frac{{\cal N}(l)}{D_j^{\nu_j}},
	\end{align}
	one can carry the similar procedure to arrive the similar form as \eqref{eq:FeyPara} with $(\prod_{j=1}^n a_j^{\nu_j-1})$ replaced by more general polynomial of $a$ and the change of powers of $\mathcal{U}$ and $\mathcal{F}$\footnote{More details of derivations can be found in \cite{Gluza:2010rn,delaCruz:2024xsm,Huang:2024nij}.}.  
	
	In this paper, we will focus on the 2-loop integrals although our discussions can be straightforwardly generalized to higher loops. For general 2-loop Feynman diagrams, the topology is just the shape of sunset and propagators can be divided into 3 branches. Following the strategy of \cite{Huang:2024nij} we label them with different parameters. For propagators of the form ${1\over -(l_1-p_{Li})^2+m_{Li}^2}$, we denote the corresponding Feynman  parameters as $(x_1,\ldots,x_{n_x})$. For propagators of the form ${1\over -(l_2-p_{Rj})^2+m_{Rj}^2}$, we denote the corresponding Feynman  parameters as $(y_{n_x+1},\ldots,y_{n_x+n_y})$. Finally for propagators of the form ${1\over -(l_1+l_2-p_{Ml})^2+m_{Ml}^2}$, we denote the corresponding Feynman  parameters as $(z_{n_x+n_y+1},\ldots,z_{n_x+n_y+n_z})$. By introducing three $\delta$-functions, $\delta(X-\sum_i x_i)$, $\delta(Y-\sum_i y_i)$, and $\delta(Z-\sum_i z_i)$, into \eqref{eq:FeyPara}, we get a new expression 
	\begin{align}
		I=&\frac{\Gamma(\nu-D)}{\prod_{j=1}^n\Gamma(\nu_j)}\int dXdYdZ ~\delta(1-X-Y-Z)\int \widetilde{d^n \boldsymbol{a}}~\mathcal{G}\,,~~~
		\label{I-form-1}
	\end{align}
	where
	\begin{align}
		\widetilde{d^n \boldsymbol{a}}&=d^n \boldsymbol{a}~ \delta(X-\sum_i x_i)\delta(Y-\sum_i y_i)\delta(Z-\sum_i z_i)\,,\notag\\
		\mathcal{G}&=(\prod_{j=1}^n a_j^{\nu_j-1})\frac{[\mathcal{U}(\boldsymbol{a})]^{\nu-\frac{3D}{2}}}{[\mathcal{F}(\boldsymbol{a})]^{\nu-D}}\,.
	\end{align}
	The explicit expressions for $\mathcal{U}$ and $\mathcal{F}$ are 
	\bea \boldsymbol{K}= \left(\begin{array}{cc} X+Z & Z\\ Z & Y+Z 
	\end{array} \right),~~~~\mathcal{U}=|\boldsymbol{K}|=XY+XZ+YZ~~~~\label{2loop-2-4}\eea
	%
	%
		%
		and 
		\bea {\mathcal F} & = & {\mathcal U}(\sum_{i=1}^{n_x} x_{i}(m_{Li}^2-p_{Li}^2)+\sum_{j=1}^{n_y} y_{j}(m_{Rj}^2-p_{Rj}^2)+
		\sum_{l=1}^{n_z} z_{l} (m_{Ml}^2-p_{Ml}^2))\nnn
		& & +(Y+Z)\sum_{i=1}^{n_x} x_{i}^2 p_{Li}^2+ (Y+Z)\sum_{1\leq i<j \leq n_x} x_{i}x_{j}(2p_{Li}\cdot p_{Lj})\nnn
		& & + (X+Z)\sum_{j=1}^{n_y} y_{j}^2p_{Rj}^2+(X+Z)\sum_{1\leq i<j \leq n_y} y_{i}y_{j}(2p_{Ri}\cdot p_{Rj})\nnn
		& & + (X+Y)\sum_{l=1}^{n_z} z_{l}^2p_{Ml}^2+(X+Y)\sum_{1\leq i<j \leq n_z} z_{i}z_{j}(2p_{Mi}\cdot p_{Mj})\nnn
		& & +(Y-Z) \sum_{i=1}^{n_x}\sum_{l=1}^{n_z} x_{i}z_{l} (p_{Li}\cdot p_{Ml})+(X-Z)\sum_{j=1}^{n_y} \sum_{l=1}^{n_z}y_{j}z_{l}(p_{Rj} \cdot p_{Ml})\nnn
		& & +(-2Z)\sum_{i=1}^{n_x}\sum_{j=1}^{n_y}x_{i}y_{j}(p_{Li}\cdot p_{Rj})~~~~~~~\label{2loop-3-3}\eea
		%
		In this paper, we will not use the explicit expression \eqref{2loop-3-3},  but only the character that ${\mathcal F}$ is a polynomial of Feynman parameters $(x,y,z)$ up to degree two. Since ${\mathcal U}$ does not depend on $x,y,z$ we can write $I$ in \eqref{I-form-1} as
		\bea
		I &  =&\frac{\Gamma(\nu-D)}{\prod_{j=1}^n\Gamma(\nu_j)}\int dXdYdZ ~\delta(1-X-Y-Z){\mathcal U}^\eta\int \widetilde{d^n \boldsymbol{a}}~{\cal P}(x,y,z)\mathcal{F}^\gamma\,,~~~
		\label{I-form-2}
		\eea
		where ${\cal P}$ is polynomial and $\eta,\gamma$ are general powers for general integrals with numerators. We will call the part $\int \widetilde{d^n \boldsymbol{a}}~{\cal P}(x,y,z)\mathcal{F}^\gamma$ as the {\bf one-loop-like integrals}, which will be the focus of the paper. We want to remark that the one-loop-like integrals is called {\bf fixed-branch integrals} in \cite{Huang:2024nij,Huang:2024qan}.
		
		For latter convenience, we write 
		\begin{align}\label{eq:Fexpr}
			\mathcal{F}&=\mathcal{C}_0+\sum_{i}\mathcal{C}_ia_i+\sum_{i,j}\mathcal{C}_{i,j}a_ia_j\notag\\
			&= \frac{1}{2}   \begin{pmatrix}
				1 & \boldsymbol{a}^T
			\end{pmatrix}
			\begin{pmatrix}
				2\mathcal{C}_0 & \boldsymbol{\mathcal{C}}^{T}\\
				\boldsymbol{\mathcal{C}} & \boldsymbol{\mathcal{A}}
			\end{pmatrix}
			\begin{pmatrix}
				1\\
				\boldsymbol{a}
			\end{pmatrix}\equiv \frac{1}{2}   \begin{pmatrix}
				1 & \boldsymbol{a}^T
			\end{pmatrix}
			{\boldsymbol{\mathcal{M}}}
			\begin{pmatrix}
				1\\
				\boldsymbol{a}
			\end{pmatrix}
		\end{align}
		where
		\begin{align}
			\boldsymbol{\mathcal{C}}_i=\mathcal{C}_i,~~~~~\boldsymbol{\mathcal{A}}_{i,j}=2 \mathcal{C}_{i,j}\,.~~~\label{AC-def}
		\end{align}
		It is worth to notice that $\boldsymbol{\mathcal{A}}$ is a symmetric matrix, i.e., $\mathcal{C}_{i,j}=\mathcal{C}_{j,i}$. When we consider the differential equation of master integrals over $\mathcal{C}$, we need to take care of this point. 
		
		\section{Complete reduction of  one-loop-like integrals}
		
		In this section, we will discuss the IBP reduction for one-loop-like integrals. First we discuss
		how to construct the IBP relations for integrands having the delta-functions. Secondly, we write down some useful IBP relations. 
		
		\subsection{IBP in the parameterization}
		
		In this part, we will discuss the reduction of $\int \widetilde{d^n \boldsymbol{a}}~{\cal P}(x,y,z)\mathcal{F}^\gamma$ in \eqref{I-form-2} for general polynomial. 
		%
		%
		To deal with the presence of delta-function, we will take a slightly different approach compared to the one in \cite{Huang:2024nij}. First using the three delta-functions, we can solve one of $x$, for example, $x_{k_1}=X-\sum_{j\neq k_1}x_j$ and similarly $y_{k_2}=Y-\sum_{j\neq k_2}y_j$ and $z_{k_3}=Z-\sum_{j\neq k_3}z_j$. After that, the $\int \widetilde{d^n \boldsymbol{a}}$ part in \eqref{I-form-2} can be written as 
		\begin{align}
			I'(\gamma,\{\nu_1,\ldots,\nu_n\})=\int \widetilde{d^n \boldsymbol{a}}~ (\prod_{j\neq k_1,k_2,k_3} a_j^{\nu_j-1})[F(\boldsymbol{a})]^{\gamma}\equiv \int \widetilde{d^n \boldsymbol{a}}~G\,,~~~\label{IBP-form-1}
		\end{align}
		where $F(\boldsymbol{k})$ is obtained from $\mathcal{F}$ by substituting $x_{k_1}$, $y_{k_2}$, and $z_{k_3}$. We will  denote it 
		\begin{align}
			F(\boldsymbol{k})&= {c}_0+\sum_{i}{c}_ia_i+\sum_{i,j}{c}_{i,j}a_ia_j=\frac{1}{2}   \begin{pmatrix}
				1 & \boldsymbol{a_{\widehat{k}}}^T
			\end{pmatrix}\begin{pmatrix}
				2c_0 & \boldsymbol{C}^{T}\\
				\boldsymbol{C} & \boldsymbol{A}
			\end{pmatrix}
			\begin{pmatrix}
				1\\
				\boldsymbol{a_{\widehat{k}}}
			\end{pmatrix}\notag\\
			&\equiv \frac{1}{2}   \begin{pmatrix}
				1 & \boldsymbol{a_{\widehat{k}}}^T\end{pmatrix}
			\boldsymbol{M}
			\begin{pmatrix}
				1\\
				\boldsymbol{a_{\widehat{k}}}
			\end{pmatrix}	\,,\label{eq:Fkexpr}
		\end{align}
		where the $\boldsymbol{\widehat{k}}$ denotes the removal of the $k_1$-th,  $k_2$-th, and $k_3$-th rows and columns. Different choices of $\boldsymbol{\widehat{k}}$ will end up different $F(\boldsymbol{k})$ from the same \eqref{eq:Fexpr}.

		Now let us consider the IBP relation of $x_i$ ($i\neq k_1$)
		\begin{align}
			T=\int_0^\infty d^n \boldsymbol{a}~ \partial_{x_i}\Big(\delta(X-\sum_j x_j)\delta(Y-\sum_j y_j)
			\delta(Z-\sum_j z_j)G\Big)\,.
		\end{align}
		On the one hand, we have 
		\begin{align}
			T =& \int \widetilde{d^{n-1} \boldsymbol{a}}~ G|_{x_{i}=\infty} -\int \widetilde{d^{n-1} \boldsymbol{a}}~ G|_{x_{i}=0}\,.
		\end{align}
		The presence of the delta function makes  the boundary term at $x_i=\infty$ vanish, leaving 
		\begin{align} 
			T & =   -\int \widetilde{d^{n-1} \boldsymbol{a}}~ G|_{x_{i}=0}=-\int \widetilde{d^{n}\boldsymbol{a}}~ \delta(x_i) G \,.
		\end{align}
		On the other hand, we have 
		\begin{align} 
			T= & \int d^{n}\boldsymbol{a}~ \Big(\partial_{x_i} \delta(X-\sum_{j} x_{j})\Big) \delta(Y-\sum_{j} y_{j})\delta(Z-\sum_{j} z_{j}) G+ \int \widetilde{d^{n}\boldsymbol{a}}~ \partial_{x_i}G\,.
		\end{align}
		Combining these results, we obtain 
		\begin{align} 
			-\int \widetilde{d^{n} \boldsymbol{a}}~ \delta(x_i) G=&  \int d^{n}\boldsymbol{a}~ \Big(\partial_{x_i}\delta(X-\sum_{j} x_{j})\Big) \delta(Y-\sum_{j} y_{j})\delta(Z-\sum_{j} z_{j}) G\notag\\
			&  + \int \widetilde{d^{n} \boldsymbol{a}}~\partial_{x_i}G\,.
		\end{align}
		Next, we consider the IBP relation of $x_{k_1}$. Following similar computations, we obtain 
		\begin{align} 
			-\int \widetilde{d^{n} \boldsymbol{a}}~ \delta(x_{k_1}) G=  \int d^{n}\boldsymbol{a}~ \Big(\partial x_{k_1}\delta(X-\sum_{j} x_{j})\Big) \delta(Y-\sum_{j} y_{j})\delta(Z-\sum_{j} z_{j}) G\,, 
		\end{align}
		where we have used the property $\partial_{x_{k_1} }G=0$ since $x_{k_1}$ has been integrated out. Now coming to the key observation 
		\begin{align} 
			\partial_{x_i}\delta(X-\sum_{j}x_{j})=\partial_{x_{k_1}} \delta(X-\sum_{j} x_{j})\,.
		\end{align}
		By subtraction we get 
		\begin{align} 
			\int  \widetilde{d^{n}\boldsymbol{a}}~  \partial_{x_{i}} G=\int  \widetilde{d^{n}\boldsymbol{a}}~ G(\delta(x_{k_1}) -\delta(x_{i}) )\,.~~~\label{IBP-form-2}
		\end{align}
		For simplicity, we can express the IBP relations as the following algebraic identities, with the understanding that these identities hold only when they are substituted back into the integration context:
		\begin{align}\label{eq:derIden-1}
			\partial_{x_{i}} G=\big[\delta(x_{k_1})-\delta(x_i)\big] G\,.
		\end{align}
		For later convenience, we will define $a_j^{I}$ to be the corresponding $a$, which appears in the same delta-function and has been integrated out. For example, $x_{k_1}$ in \eqref{eq:derIden-1} will be denoted as $x_i^I$. Using this notation, we can rewrite \eqref{eq:derIden-1} as 
		\bea \partial_{a_{i}} G= \O{\delta}(a_i) G,~~~~~~\O{\delta}(a_i)\equiv \delta(a_i^I)-\delta(a_i)\label{eq:derIden}\eea
		Noticing that although originally there are $n$ Feynman parameters, there are only $(n-3)$ IBP relations in the form  \eqref{eq:derIden}.
		
		\subsection{Iterative IBP reduction relations}
		
		In this part, we will consider the IBP relation of the form
		\bea \sum_i \frac{\partial (P_i(a) QF^{\gamma})}{\partial a_i}=\sum_i \O{\delta}(a_i)P_i(a) QF^{\gamma}~~~~\label{4.1}\eea
		on the left-hand side of \eqref{eq:derIden}, where $F$ is given in \eqref{eq:Fkexpr}, $P(a)$ are polynomials of $a$ and 
		\bea Q(\vec{\nu})=\prod_i a_i^{\nu_i}\,.~~~~\label{Q-def}\eea
		With some nice choices of $P_i(a)$, we will get several useful IBP relations. Using these relations, we can find the master integrals and obtain reduction coefficients of any integrals to these master integrals. Using these results, we can derive the differential equations of master integrals in the next section.
		
		\subsubsection{The first type of choices}
		Now let us consider the first type of choice. Fixing an index $i_0$, we take 
		\bea P_{i_0}=F,~~~P_{j\neq i_0}=0~~~~\label{syz-T2-1}\eea
		Putting it back to \eqref{4.1}, it gives 
		\bea {\nu_{i_0}\over a_{i_0}} Q(\vec{\nu})  F^{\gamma+1}
		+(\gamma+1) {\partial  F \over \partial a_{i_0}} Q(\vec{\nu}) F^{\gamma}=\O{\delta}(a_{i_0}) Q(\vec{\nu})  F^{\gamma+1}~~~~\label{syz-T2-2}\eea
		Let us define following useful action\footnote{It is important to notice that ${\bm i}^\pm$ act only on $Q(\vec{\nu})$, not on $F$.} 
		\bea {\bm i}_0^- Q(\vec{\nu})\equiv {\partial \over \partial a_i } Q(\vec{\nu})={\nu_{i_0}\over a_{i_0}}Q(\vec{\nu}),~~~~~~~{\bm i}_0^+ Q(\vec{\nu})\equiv  a_{i_0}Q(\vec{\nu}) ~~~\label{am-action}\eea
		Using them, \eqref{syz-T2-2} can be written as 
		\bea {\bm i}_0^- Q(\vec{\nu})  F^{\gamma+1}
		+(\gamma+1)(c_{i_0}+2\sum_{j} c_{i_0 j} \bm{j}^+)  Q(\vec{\nu}) F^{\gamma}=\O{\delta}(a_{i_0}) Q(\vec{\nu})  F^{\gamma+1}~~~~\label{Type-1-IBP}\eea
		The relation \eqref{Type-1-IBP} contains different powers of $F$, so it is like the relation connecting different dimensions. It will be useful when we consider the differential equation for master integrals. To see its usefulness, 
		let us consider several applications of the formula:
		\begin{itemize}
			\item[(a)] When taking $Q(\vec{\nu})=1$ and $i_0=1,...,\O n$ (for simplicity, we have written $\O n=n-3$), \eqref{Type-1-IBP} can be written in the  matrix form 
			\bea \boldsymbol{A}\cdot \boldsymbol{a} F^{\gamma}= -\boldsymbol{C} F^{\gamma}+{1\over (\gamma+1)}\boldsymbol{\O\delta} F^{\gamma+1}~~~\label{T1-1-1}\eea
			where $\boldsymbol{A},\boldsymbol{C}$ are given in \eqref{eq:Fkexpr} (especially, $A_{ij}=2 c_{ij}$) and  $\boldsymbol{a},\boldsymbol{\O\delta}$ are row vectors
			\bea \boldsymbol{a}^T=(a_1,...,a_{\O n});~~~~~~\boldsymbol{\O\delta}^T=(\O\delta(a_1),...,\O\delta(a_{\O n}))~~~\label{T1-1-2}\eea
			If the matrix $\boldsymbol{A}$ is non-degenerate, we can reduce the rank one integrals\footnote{Here we define the rank of the integrals $Q(\vec{\nu})  F^{\gamma+1}$ to the $|\nu|\equiv \sum_i \nu_i$. Also, we call
				$\delta(a_i)F^{\gamma}$ to be the subsector of $F^{\gamma}$ since the number of Feynman parameters is reduced by one.} to the scalar basis and subsectors
			\bea \boldsymbol{a} F^{\gamma}= \boldsymbol{A}^{-1}\cdot\left\{-\boldsymbol{C} F^{\gamma}+{1\over (\gamma+1)}\boldsymbol{\O\delta} F^{\gamma+1}\right\}~~~\label{T1-1-3}\eea
			If matrix $\boldsymbol{A}$ is degenerate, \eqref{T1-1-1} means the scalar basis $F^{\gamma}$ is not a basis anymore and can be reduced to subsectors. One way to see it is that now there is an eigenvector of $\boldsymbol{A}$ with eigenvalue zero, i.e., $\boldsymbol{\alpha}\cdot\boldsymbol{A}=0$. Multiplying at both sides we get\footnote{There may be more than one eigenvector with eigenvalue zero; different choices of $\boldsymbol{\alpha}$ will lead to extra relations between ${\O\delta}^{T} F^{\gamma+1}$, which means subsectors are not all independent, i.e., some subsectors will not be master integrals anymore.}
			\bea  F^{\gamma}={1\over (\gamma+1)\boldsymbol{\alpha}\cdot\boldsymbol{C}^{T} }\boldsymbol{\alpha}\cdot\boldsymbol{\O\delta} F^{\gamma+1}~~~\label{T1-1-4}\eea

			\item[(b)] For the $Q(\vec{\nu})$ with rank one, we can write $Q(\vec{\nu})=\boldsymbol{a}^T$ as a row vector. 
			Using $\boldsymbol{i}^-\cdot \boldsymbol{a}^T=I_{\O n\times\O n}$, \eqref{Type-1-IBP} becomes the matrix form 
			\bea I_{\O n\times\O n}  F^{\gamma+1}
			+(\gamma+1)(\boldsymbol{C} +\boldsymbol{A}\cdot \boldsymbol{a}) \cdot \boldsymbol{a}^T F^{\gamma}=\boldsymbol{\O\delta}\cdot  \boldsymbol{a}^T F^{\gamma+1}~~~\label{T1-2-1}\eea
			If we take the trace at the both side, we will get 
			\bean\O  n F^{\gamma+1}
			+(\gamma+1)(2F-\boldsymbol{C}^T\cdot \boldsymbol{a}-2c_0)  F^{\gamma}=\left\{\delta(x_{k_1})X+\delta(y_{k_2})Y+\delta(z_{k_3})Z \right\} F^{\gamma+1}\eean
			where we have used the result 
			\bea & & \int  \widetilde{d^{n}\boldsymbol{a}}~\text{Tr}(\boldsymbol{\O\delta}\cdot  \boldsymbol{a}^T)=\int  \widetilde{d^{n}\boldsymbol{a}}~\boldsymbol{\O\delta}^T\cdot  \boldsymbol{a}=\int  \widetilde{d^{n}\boldsymbol{a}}
			\sum_i (\delta(a_i^I)-\delta(a_i)) a_i \nnn
			& = & \int  \widetilde{d^{n}\boldsymbol{a}}
			\sum_i \delta(a_i^I)a_i=\int  \widetilde{d^{n}\boldsymbol{a}}\left\{\delta(x_{k_1})X+\delta(y_{k_2})Y+\delta(z_{k_3})Z \right\}~~~\label{T1-2-3} \eea
			Rearranging it we can write 
			\bea (\O n+2(\gamma+1))F^{\gamma+1} & = & (\gamma+1)(\boldsymbol{C}\cdot \boldsymbol{a}^T+2c_0)  F^{\gamma} +\delta_{XYZ} F^{\gamma+1}~~~\label{T1-2-4} \eea
			where for simplification, we have defined
			\bea \delta_{XYZ}\equiv \delta(x_{k_1})X+\delta(y_{k_2})Y+\delta(z_{k_3})Z~~~\label{T1-2-5-1} \eea
			If the matrix $\boldsymbol{A}$ is non-degenerate, we can use \eqref{T1-1-3} to simplify \eqref{T1-2-4} further as 
			\bea  (\O n+2(\gamma+1))F^{\gamma+1} &=& \left\{ 2(\gamma+1)c_0  -{(\gamma+1)} \boldsymbol{C}^T\cdot \boldsymbol{A}^{-1}\cdot\boldsymbol{C}\right\}F^{\gamma}\nnn & & +\left\{  \boldsymbol{C}^T\cdot \boldsymbol{A}^{-1}\cdot\boldsymbol{\O\delta}  +\delta_{XYZ} \right\} F^{\gamma+1}~~~\label{T1-2-5} \eea
			%
			%
			Now we can consider the reduction of rank two tensor using \eqref{T1-2-1}. Rewritting it as 
			\bea 
			\boldsymbol{A}\cdot \boldsymbol{a} \cdot \boldsymbol{a}^T F^{\gamma}=-\boldsymbol{C} \cdot \boldsymbol{a}^T F^{\gamma}+{1\over (\gamma+1)}\left\{-I_{\O n\times\O n}  F^{\gamma+1}+\boldsymbol{\O\delta}\cdot  \boldsymbol{a}^T F^{\gamma+1}\right\}~~~\label{T1-2-6}\eea
			we can solve 
			\bea 
			\boldsymbol{a} \cdot \boldsymbol{a}^T F^{\gamma}=-\boldsymbol{A}^{-1}\cdot\boldsymbol{C} \cdot \boldsymbol{a}^T F^{\gamma}+{1\over (\gamma+1)}\boldsymbol{A}^{-1}\cdot\left\{-I_{\O n\times\O n}  F^{\gamma+1}+\boldsymbol{\O\delta}\cdot  \boldsymbol{a}^T F^{\gamma+1}\right\}~~~~~\label{T1-2-7}\eea
			when the matrix $\boldsymbol{A}$ is non-degenerate. The $F^{\gamma+1}$ at the right-hand side can be simplified further using  \eqref{T1-2-5}. If $\boldsymbol{A}$ is  degenerate and there is zero eigenvector such that $\boldsymbol{\alpha}^T\cdot \boldsymbol{C}  \neq 0$, we can solve 
			\bea \boldsymbol{a}^T F^{\gamma}={1\over (\gamma+1)\boldsymbol{\alpha}^T\cdot \boldsymbol{C}}\boldsymbol{\alpha}^T\cdot \left\{-I_{\O n\times\O n}  F^{\gamma+1}+\boldsymbol{\O\delta}\cdot  \boldsymbol{a}^T F^{\gamma+1}\right\}~~~~~\label{T1-2-8}\eea
			which reduces the rank one tensor integrals.
		\end{itemize}
		
		\subsubsection{The second type of choices}
		
		For this one, we take 
		\bea P_{i\neq i_0}(\boldsymbol{a})= a_i {\partial F(\boldsymbol{a}) \over \partial a_{i_0}},~~~~P_{i_0}(\boldsymbol{a})= a_{i_0} {\partial F(\boldsymbol{a}) \over \partial a_{i_0}}+2 c_0 +\sum_{j} c_{j} a_{j},~~~~~~~~\label{syz-T3-1}  \eea
		Putting it to \eqref{4.1} and doing some algebraic simplifications, we can get 
		\bea & & \left\{ (2c_0+\sum_j c_j \bm{j}^+ )\bm{i}_0^- +(2{\gamma}+\O n+1+\sum_{j}\bm{j}^+\bm{j}^-)(c_{i_0}+2\sum_{j} c_{i_0 j} \bm{j}^+)\right\} Q(\vec{\nu})F^{{\gamma}}\nnn
		& = & \left\{ \delta_{XYZ}(c_{i_0}+2\sum_{j} c_{i_0 j} \bm{j}^+)  +\O\delta(a_{i_0})(2 c_0 +\sum_{j} c_{j} \bm{j}^+) \right\} Q(\vec{\nu})F^{{\gamma}}~~~~~~~~~\label{Type-2-IBP-1}\eea
		Defining $\nu_{Q}=\sum_i \nu_i$ for $Q(\vec{\nu})$ and rewriting \eqref{Type-2-IBP-1} to matrix form, we have 
		\bea & & (2{\gamma}+\O n+1+\nu_{Q}+1)\boldsymbol{A}\cdot \boldsymbol{a} Q(\vec{\nu})F^{{\gamma}}\nnn
		& = & -(2{\gamma}+\O n+1+\nu_{Q}) \boldsymbol{C} Q(\vec{\nu})F^{{\gamma}} - (2c_0+\boldsymbol{C}^T\cdot \boldsymbol{a} )(\boldsymbol{i}^{-})Q(\vec{\nu})F^{{\gamma}}\nnn
		& & +\left\{ \delta_{XYZ}(\boldsymbol{C}+\boldsymbol{A}\cdot \boldsymbol{a})  +\boldsymbol{\O\delta}(2c_0+\boldsymbol{C}^T\cdot \boldsymbol{a} ) \right\} Q(\vec{\nu})F^{{\gamma}}~~~~~~~~~\label{Type-2-IBP-2} \eea
		A good point comparing to \eqref{Type-1-IBP} is that the power of $F$ is the same in \eqref{Type-2-IBP-2}. 
		When $|\boldsymbol{A}|\neq 0$, we can use this formula to reduce the higher rank tensor integrals on the left-hand side to the lower rank tensor integrals as well as the subsectors on the right-hand side. When  $|\boldsymbol{A}|= 0$, we can also use it to reduce some higher rank tensor integrals. To add extra relations to do the reduction, we can use the following strategy. Taking the zero eigenvector $\boldsymbol{\alpha}_i$ of $\boldsymbol{A}$, we multiply it at the both side and rewrite $Q(\vec{\nu})\to \boldsymbol{a} Q(\vec{\nu})$ to get
		\bea 
		&  & (2{\gamma}+\O n+1+\nu_{Q}+1)\boldsymbol{\alpha}_i^T\cdot \boldsymbol{C} \boldsymbol{a} Q(\vec{\nu})F^{{\gamma}} + (2c_0+\boldsymbol{C}^T\cdot \boldsymbol{a} )\boldsymbol{\alpha}_i^T\cdot (\boldsymbol{i}^{-}) \boldsymbol{a} Q(\vec{\nu})F^{{\gamma}}\nnn
		& =& \boldsymbol{\alpha}_i^T\cdot \left\{ \delta_{XYZ}(\boldsymbol{C}+\boldsymbol{A}\cdot \boldsymbol{a})  +\boldsymbol{\O\delta}(2c_0+\boldsymbol{C}^T\cdot \boldsymbol{a} ) \right\} \boldsymbol{a} Q(\vec{\nu})F^{{\gamma}}~~~~~~~~~\label{Type-2-IBP-3} \eea
		Taking all $\boldsymbol{\alpha}_i$ we get the wanted extra relations to give the full reduction when combining with \eqref{Type-2-IBP-2}.
		
		Now we present some simple applications of \eqref{Type-2-IBP-2}:
		\begin{itemize}
			\item[(a)] When $Q(\vec{\nu})=1$, we get
			\bea & & (2{\gamma}+\O n+2)\boldsymbol{A}\cdot \boldsymbol{a} F^{{\gamma}}= -(2{\gamma}+\O n+1) \boldsymbol{C} F^{{\gamma}} \nnn
			& & +\left\{ \delta_{XYZ}(\boldsymbol{C}+\boldsymbol{A}\cdot \boldsymbol{a})  +(2c_0+\boldsymbol{C}^T\cdot \boldsymbol{a} ) \boldsymbol{\O\delta}\right\} F^{{\gamma}}~~~~~~~~~\label{T2-1-1} \eea
			It is similar to \eqref{T1-1-1}. When combining them together, we can have extra relations to reduce $\boldsymbol{\O\delta}F^{{\gamma+1}}$.
			
			\item[(b)] When $Q(\vec{\nu})=\boldsymbol{a}$, we get
			\bea & & (2{\gamma}+\O n+3)\boldsymbol{A}\cdot (\boldsymbol{a} \cdot \boldsymbol{a}^T)F^{{\gamma}}\nnn
			& = & -(2{\gamma}+\O n+2) \boldsymbol{C} \cdot \boldsymbol{a}^TF^{{\gamma}} - (2c_0+\boldsymbol{C}^T\cdot \boldsymbol{a} )I_{\O n\times\O n}F^{{\gamma}}\nnn
			& & +\left\{ \delta_{XYZ}(\boldsymbol{C}+\boldsymbol{A}\cdot \boldsymbol{a})  +(2c_0+\boldsymbol{C}^T\cdot \boldsymbol{a} ) \boldsymbol{\O\delta}\right\} \cdot \boldsymbol{a}^TF^{{\gamma}}~~~~~~~~~\label{T2-2-1} \eea
			which can reduce the rank two tensor integrals. 
		\end{itemize}
		
		\subsubsection{The third type of choices}
		For this one, we take 
		\bea P_i(\boldsymbol{a})=a_i,~~~~~~~~\label{syz-T4-1}  \eea
		The IBP relation is 
		\bea & &\left(\sum_i \bm{i}^+\bm{i}^-+\O n+2{\gamma}\right)Q(\vec{\nu})  F^{{\gamma}}-{\gamma} (2c_0+\sum_j c_j \bm{j}^+  ) Q(\vec{\nu})F^{{\gamma}-1}\nnn &=& \left(\sum_i \O\delta(a_i)a_i\right) Q(\vec{\nu})  F^{{\gamma}}~~~~~~~~~\label{syz-T4-3}\eea
		Using the result \eqref{T1-2-3}, it becomes
		\bea & &\left(\sum_i \bm{i}^+\bm{i}^-+\O n+2{\gamma}\right)Q(\vec{\nu})  F^{{\gamma}}-{\gamma} (2c_0+\sum_j c_j \bm{j}^+  ) Q(\vec{\nu})F^{{\gamma}-1}=\delta_{XYZ} Q(\vec{\nu})  F^{{\gamma}}~~~~~~~~~\label{Type-3-IBP}\eea
		Again, we see a few examples:
		\begin{itemize}
			\item[(a)] When $Q(\vec{\nu})=1$, we have 
			\bea & &\left(\O n+2{\gamma}\right) F^{{\gamma}}-{\gamma} (2c_0+\boldsymbol{C}^T\cdot \boldsymbol{a} ) F^{{\gamma}-1}=\delta_{XYZ}   F^{{\gamma}}~~~~~~~~~\label{T3-1-1}\eea
			It is nothing, but the one \eqref{T1-2-4}.
			
			\item[(b)] When $Q(\vec{\nu})=\boldsymbol{a}$, we get
			\bea & &\left(1+\O n+2{\gamma}\right)\boldsymbol{a}  F^{{\gamma}}-{\gamma} (2c_0+\boldsymbol{C}^T\cdot \boldsymbol{a} ) \boldsymbol{a}F^{{\gamma}-1}=\delta_{XYZ} \boldsymbol{a} F^{{\gamma}}~~~~~~~~~\label{T3-2-1}\eea
		\end{itemize}

		\section{\label{sec:des}Master integrals and their differential equations}
		
		From above discussions, one can see that any integral $I'(\gamma,\{\nu_1,\ldots,\nu_n\})$ can be reduced to integrals
		$I'(\gamma,\boldsymbol{w})\equiv I'(\gamma,\{w_1,\ldots,w_n\})$ with $w_i=0,1$. It is similar to the fact that any one-loop integral can be reduced to the scalar integrals.
		With this observation, the first natural choice is to take the master integrals to be $I'(\gamma+1,\boldsymbol{w})$.
		
		Now let us see the differential equations for these chosen master integrals. For $\frac{\partial I'(\gamma+1,\boldsymbol{w})}{\partial c_0}$, we have\footnote{Again, equation \eqref{F-c0-1}
			holds only under the integration. However, since ${\d\over \d c_0}$ commutes with integration, we can write it as the algebraic relation. },
		\bea \frac{\partial F^{\gamma+1}}{\partial c_0}=(\gamma+1)F^{\gamma}& =&-\frac{2\gamma+n-1}{{\mathcal{D}}} F^{\gamma+1}+\frac{\boldsymbol{C}^T \cdot \boldsymbol{A}^{-1}\cdot \boldsymbol{\O\delta}+\delta_{XYZ}}{\mathcal{D}}    F^{\gamma+1}~~~\label{F-c0-1}\eea
		where the second equation has used the result \eqref{T1-2-5} and 
		\begin{align}
			\mathcal{D}&=\boldsymbol{C}^T\cdot\boldsymbol{A}^{-1}\cdot \boldsymbol{C} -2c_0\,. ~~~\label{cal-D}
		\end{align}
		Here and in the later part of this section, we have assumed $\boldsymbol{A}$ is non-degenerate. The result \eqref{F-c0-1} is nice since a given master integral in the differential equation depends only on itself and the nearest subsector.  
		
		Now we consider the differential equations for $\frac{\partial F^{\gamma+1}}{\partial c_i}$. 
		Similarly we have 
		\bea \frac{\partial F^{\gamma+1}}{\partial c_i}=(\gamma+1) a_i F^{\gamma}& =&-(\gamma+1)(\boldsymbol{A}^{-1}\cdot \boldsymbol{C})_i F^{\gamma}+(\boldsymbol{A}^{-1}\cdot \boldsymbol{\O \delta})_i F^{\gamma+1}~~~\label{F-ci-1}\eea
		where \eqref{T1-1-3} has been used. Using \eqref{F-c0-1} again, we get 
		\bea & &\frac{\partial F^{\gamma+1}}{\partial c_i}=(\gamma+1) a_i F^{\gamma}\nnn
		& =&-(\boldsymbol{A}^{-1}\cdot \boldsymbol{C})_i \left( -\frac{2\gamma+n-1}{{\mathcal{D}}} F^{\gamma+1}+\frac{\boldsymbol{C}^T \cdot \boldsymbol{A}^{-1}\cdot \boldsymbol{\O\delta}+\delta_{XYZ}}{\mathcal{D}}    F^{\gamma+1}\right)\nnn & &+(\boldsymbol{A}^{-1}\cdot \boldsymbol{\O \delta})_i F^{\gamma+1}~~~\label{F-ci-2}\eea
		Again we see a given master integral in the differential equation depends only on itself and the nearest subsector. 

		Finally we consider the differential equations for $\frac{\partial F^{\gamma+1}}{\partial c_{i,j}}$\footnote{When we take the derivative over $c_{ij}$ in \eqref{F-cij-1}, we have assumed
			$c_{ij}$ to be independent. If one insistent the condition $c_{ij}=c_{ji}$, one just need to add up $\frac{\partial F^{\gamma+1}}{\partial c_{i,j}}$ and $\frac{\partial F^{\gamma+1}}{\partial c_{ji}}$. }. We have 
		\bea & & \frac{\partial F^{\gamma+1}}{\partial c_{i,j}}= (\gamma+1) a_i a_j F^{\gamma}\nnn &=&-(\gamma+1)(\boldsymbol{A}^{-1}\cdot\boldsymbol{C})_i a_j F^{\gamma}-(\boldsymbol{A}^{-1})_{ij} F^{\gamma+1}+(\boldsymbol{A}^{-1}\cdot\boldsymbol{\O\delta})_i a_j F^{\gamma+1},~~~\label{F-cij-1}\eea
		where \eqref{T1-2-7} has been used. To reduce to the chosen master integrals, the first term in the second line of \eqref{F-cij-1} should be replaced by the result in \eqref{F-ci-2} and we see the dependence of itself and the nearest subsector.  The trouble part is the third term $a_j F^{\gamma+1}$ of the nearest subsector. Reducing this term will produce dependence in all subsectors, which makes the pattern of differential equation complicated.  There is another disadvantage, i.e., the differential equations \eqref{F-c0-1} and \eqref{F-ci-2} are not canonical. 
		
		Now we want to construct the canonical basis from the natural scalar basis, i.e., looking at the basis of the form $g(c) F^{\gamma+1}$.  
		Considering the action $\partial_{c_{ij}}$, we have 
		\bea \frac{\partial g(c)F^{\gamma+1}}{\partial c_{i,j}}& = & \frac{\partial g(c)}{\partial c_{i,j}} F^{\gamma+1}-(\boldsymbol{A}^{-1})_{ij} g(c) F^{\gamma+1}\nnn & & -(\gamma+1)(\boldsymbol{A}^{-1}\cdot\boldsymbol{C})_i a_j g(c)F^{\gamma}+(\boldsymbol{A}^{-1}\cdot\boldsymbol{\O\delta})_i a_j g(c) F^{\gamma+1},~~~\label{F-cij-3}\eea
		We can  eliminate the first line by demanding 
		\bea\label{eq:elimateAij}
		0=\frac{\partial g}{\partial c_{i,j}}-g(\boldsymbol{A}^{-1})_{i,j}=\frac{\partial g}{\partial c_{i,j}}-g \frac{\partial_{\boldsymbol{A}_{i,j}}|{\boldsymbol{A}}|}{|\boldsymbol{A}|}=\frac{\partial g}{\partial c_{i,j}}-g \frac{ \partial c_{i,j}}{\partial \boldsymbol{A}_{i,j}}\frac{\partial_{c_{i,j}}|{\boldsymbol{A}}|}{|\boldsymbol{A}|}\,.
		\eea
		Using $\boldsymbol{A}_{i,j}=2c_{i,j}$, the solution is 
		\begin{align}
			g=\O g(c_0,c_i)|\boldsymbol{A}|^{1/2}\,.\label{g-1}
		\end{align}
		where $\O g$ depends only on $c_0,c_i$. 
		
		Next, we want the appearance of the overall $\epsilon$ factor in the differential equation.  To get the hind, let us look equation \eqref{F-c0-1}. For scalar integrals, from \eqref{eq:FeyPara}, we can see that
		\bea \gamma={LD\over 2}-n={L(d-2\epsilon)\over 2}-n\eea
		where $n$ is the number of propagators of this sector. Thus
		\bea 2\gamma+n-1=2(d-2\epsilon)-n-1=2d-n-1-4\epsilon\eea
		where we have set $L=2$ for our case. When $n$ is odd, we can take
		the space-time dimension to be $d={n+1\over 2}$ and $2\gamma+n-1=-4\epsilon$. However, when $n$ is even, we can only take $d={n\over 2}$ and now $2\gamma+n-1=-1-4\epsilon$. To cure this point, we need another nontrivial factor $\O g(c)$ in the definition of canonical basis in \eqref{g-1}. With the above explanation, now we define the canonical basis:
		\bea \mathcal{I}_{2m}&=& \epsilon^{m}\Gamma(m-1+2\epsilon)|\boldsymbol{A}|^{1/2}\mathcal{D}^{1/2}F_{2m}^{1-m-2\epsilon}\label{Ieven}\\
		\mathcal{I}_{2m+1}&= &\epsilon^{m+1}\Gamma(m-1+2\epsilon)|\boldsymbol{A}|^{1/2}F_{2m+1}^{1-m-2\epsilon}\label{Iodd}\eea
		Here the subscript gives the number of Feynman parameters before integrating out three $a$ using the delta-functions, thus for odd case, $m\geq 1$ and for even case $m\geq 2$.

		Now we present canonical differential equation  according to the value of $n$.
		
		\subsection{The case of $n=2m$}

		By combining \eqref{F-c0-1}, \eqref{F-ci-1}
		and \eqref{F-cij-1}, we have 
		\bea d{\cal I}_{2m} & = & c_{2m\to 2m}{\cal I}_{2m}+\sum_i c_{2m\to 2m-1;i}  {\cal I}_{2m-1}^{(i)}+\sum_{i\neq j} c_{2m\to 2m-2;ij}  {\cal I}_{2m-2}^{(ij)}\label{In-diff-1}\eea
		In \eqref{In-diff-1} the summation of $i$ is $i=1,...,n$, where the three integrated indices should also be included. For the summation $\sum_{i\neq j}$ similar understanding should be taken. An important observation of \eqref{In-diff-1} is that the right-hand side is up to sub-sub-sectors only.  Now we give the expressions of coefficients $c$:
		\begin{itemize}
			\item[(a)] For $c_{2m\to 2m}$, it is easy to find 
			\bea c_{2m\to 2m}&=& \frac{4\epsilon}{\mathcal{D}}\left\{ dc_0-(\boldsymbol{A}^{-1}\boldsymbol{C})_i dc_{i}-(\boldsymbol{A}^{-1}\boldsymbol{C})_i(\boldsymbol{A}^{-1}\boldsymbol{C})_jdc_{i,j}\right\}\nnn
			&= & -2\epsilon d\log\mathcal{D}\label{c2mto2m}\eea
			\item[(b)] For the coefficients $c_{2m\to 2m-1;i}$, using \eqref{F-c0-1}, \eqref{F-ci-1}
			and \eqref{F-cij-1} we will get the combination $|\boldsymbol{A}_{\widehat{i}}|^{1/2}\delta(a_i)F_{2m}^{2-m-2\epsilon}$
			where $\boldsymbol{A}_{\widehat{i}}$ is the matrix obtained from  $\boldsymbol{A}$ by removing the $i$-th row and column.
			However, $\delta(a_i)F_{2m}^{2-m-2\epsilon}=F_{2m-1}^{2-m-2\epsilon}$ which gives the basis \eqref{Iodd}. With this 
			clarification, we find that when $i\notin(k_1,k_2,k_3)$,
			\bea c_{2m\to 2m-1;i}  & = & \frac{-2\epsilon(\boldsymbol{C}^T \boldsymbol{A}^{-1})_i|\boldsymbol{A}|^{1/2} }{\sqrt{(\mathcal{D}-\mathcal{D}_{\widehat{i}})}|\boldsymbol{A}_{\widehat{i}}|^{1/2}}d\log\left(\frac{(\boldsymbol{C}^T \boldsymbol{A}^{-1})_i|\boldsymbol{A}|^{1/2}/|\boldsymbol{A}_{\widehat{i}}|^{1/2}-\sqrt{\mathcal{D}}}{(\boldsymbol{C}^T \boldsymbol{A}^{-1})_i|\boldsymbol{A}|^{1/2}/|\boldsymbol{A}_{\widehat{i}}|^{1/2}+\sqrt{\mathcal{D}}}\right)\nnn
			& = & -2\epsilon d\log\left(\frac{(\boldsymbol{C}^T \boldsymbol{A}^{-1})_i|\boldsymbol{A}|^{1/2}/|\boldsymbol{A}_{\widehat{i}}|^{1/2}-\sqrt{\mathcal{D}}}{(\boldsymbol{C}^T \boldsymbol{A}^{-1})_i|\boldsymbol{A}|^{1/2}/|\boldsymbol{A}_{\widehat{i}}|^{1/2}+\sqrt{\mathcal{D}}}\right)\label{eq:n2n-1}\eea
			where $\mathcal{D}_{\widehat{i}}=\boldsymbol{C}_{\widehat{i}}^T\cdot(\boldsymbol{A}_{\widehat{i}})^{-1}\cdot \boldsymbol{C}_{\widehat{i}} -2c_0$. Later we will meet $\boldsymbol{A}_{\widehat{ij}}$ and $\mathcal{D}_{\widehat{ij}}$ and the
			similar understanding should be taken. 
			The proof from the first line to second line is as follows. Take $i=1$ as an example, expanding   $\mathcal{D}-\mathcal{D}_{\widehat{1}}$ we will get 
			\bean T=c_1^2 \boldsymbol{A}^{-1}_{1,1}+2c_{1}\sum_{j\neq 1}\boldsymbol{A}^{-1}_{1,j}c_j+\sum_{k,k\neq 1}\big[c_k^2(\boldsymbol{A}^{-1}_{k,k}-\boldsymbol{A}^{-1}_{\widehat{1};k,k})+2c_{k}\sum_{j\neq 1,k}(\boldsymbol{A}^{-1}_{k,j}-\boldsymbol{A}^{-1}_{\widehat{1};k,j})c_j\big]\eean
			To continue, we need to use the Jacobi's  identity (see \cite{CARACCIOLO2013474} Lemma A.1 (e))
			\bea |\boldsymbol{A}^I_J|=(-)^{I+J}|\boldsymbol{A}||(\boldsymbol{A}^{-1})^{\O I}_{\O J}|\label{Jacobi}\eea
			where $\boldsymbol{A}^I_J$ is the submatrix constructed from the elements $A_{ij}, i\in I, j\in J$ and $\boldsymbol{A}^{\O I}_{\O J}$ is the submatrix constructed from the elements $A_{ij}, i\not\in I, j\not\in J$. The factor $(-)^{I+J}$ is $(-)^{\sum_{i\in I}i+\sum_{j\in J}j}$. Using \eqref{Jacobi}, we have 
			\bea \boldsymbol{A}^{-1}_{\widehat{1};k,j}&= & \frac{\left(\text{adj}\boldsymbol{A}_{\widehat{1}}\right)_{k,j}}{|\boldsymbol{A}_{\widehat{1}}|
			}= {(-)^{k+j}|\boldsymbol{A}^{\O{(1k)}}_{\O{(1j)}}|\over |\boldsymbol{A}_{\widehat{1}}|}\nnn
			&= &{|\boldsymbol{A}||(\boldsymbol{A}^{-1})^{1k}_{1j}|\over |\boldsymbol{A}_{\widehat{1}}|}=
			\frac{|\boldsymbol{A}|(\boldsymbol{A}^{-1}_{1,1}\boldsymbol{A}^{-1}_{k,j}-\boldsymbol{A}^{-1}_{1,k}\boldsymbol{A}^{-1}_{1,j})}{|\boldsymbol{A}|\boldsymbol{A}^{-1}_{1,1}} \eea
			thus
			\begin{align}\label{eq:pref}
				T=&c_1^2 \boldsymbol{A}^{-1}_{1,1}+2c_{1}\sum_{j\neq 1}\boldsymbol{A}^{-1}_{1,j}c_j+\sum_{k,k\neq 1}\big[c_k^2(\boldsymbol{A}^{-1}_{k,k}-\frac{|\boldsymbol{A}|(\boldsymbol{A}^{-1}_{1,1}\boldsymbol{A}^{-1}_{k,k}-(\boldsymbol{A}^{-1}_{1,k})^2)}{|\boldsymbol{A}|\boldsymbol{A}^{-1}_{1,1}})\notag\\
				&+2c_{k}\sum_{j\neq 1,k}(\boldsymbol{A}^{-1}_{k,j}-\frac{|\boldsymbol{A}|(\boldsymbol{A}^{-1}_{1,1}\boldsymbol{A}^{-1}_{k,j}-\boldsymbol{A}^{-1}_{1,k}\boldsymbol{A}^{-1}_{1,j})}{|\boldsymbol{A}|\boldsymbol{A}^{-1}_{1,1}})c_j\big]\notag\\
				=&\frac{((\boldsymbol{C}^T \boldsymbol{A}^{-1})_1)^2|\boldsymbol{A}|}{|\boldsymbol{A}_{\widehat{1}}|}\,.
			\end{align}
			
			When $i\in(k_1,k_2,k_3)$, for example, $i=k_1$, we find 
			\bea\label{eq:n2n-1prime}
			c_{2m\to 2m-1;k_1}  & = &-\frac{2\epsilon|\boldsymbol{A}|^{1/2}\sum_{i=1,i\neq k_1}^{n_x}(\boldsymbol{C}^T \boldsymbol{A}^{-1})_i }{\sqrt{(\mathcal{D}-\mathcal{D}')}|\boldsymbol{A}'|^{1/2}}d\log\left(\frac{\sqrt{\mathcal{D}-\mathcal{D}'}-\sqrt{\mathcal{D}}}{\sqrt{\mathcal{D}-\mathcal{D}'}+\sqrt{\mathcal{D}}}\right)\,,\nnn
			&=& -2\epsilon d\log\left(\frac{\sqrt{\mathcal{D}-\mathcal{D}'}-\sqrt{\mathcal{D}}}{\sqrt{\mathcal{D}-\mathcal{D}'}+\sqrt{\mathcal{D}}}\right)
			\eea
			The $\mathcal{D}'$ and $\boldsymbol{A}'$ are obtained as follows. First we replace $x_{k_1'}=X-\sum_{i=1,i\neq k_1,k_1'}^{n_x}$ in the
			$F$ (see \eqref{eq:Fkexpr}) and get the new $c_0',\boldsymbol{A}', \boldsymbol{C}' $. Then we construct $\mathcal{D}'$ using 
			\eqref{cal-D}. In the appendix, we will prove that \eqref{eq:n2n-1prime} can be expressed in the same form as \eqref{eq:n2n-1}
			with the understanding that now $c_0,\boldsymbol{A}, \boldsymbol{C} $ are read out from the $F$, which is obtained from $\mathcal{F}$ (see \eqref{eq:Fexpr}) by integrating out $x_{k_1'}$ instead of $x_{k_1}$.

			\item[(c)] For coefficients $c_{2m\to 2m-2;ij}$, we again encounter the combination $\delta(a_i)\delta(a_j)F_{2m}^{2-m-2\epsilon}$ $=F_{2m-2}^{1-(m-1)-2\epsilon}$, which is the basis \eqref{Ieven}. 
			When $i\notin(k_1,k_2,k_3)$ and $j\notin(k_1,k_2,k_3)$, direct computation gives 
			\begin{align}\label{eq:n2n-2}	
				c_{2m\to 2m-2;ij}= \frac{-\epsilon N}{2}d\log\left(\frac{\sqrt{(\mathcal{D}_{\widehat{i}}-\mathcal{D})\mathcal{D}_{\widehat{i,j}}}-\sqrt{(\mathcal{D}_{\widehat{i}}-\mathcal{D}_{\widehat{i,j}})\mathcal{D}}}{\sqrt{(\mathcal{D}_{\widehat{i}}-\mathcal{D})\mathcal{D}_{\widehat{i,j}}}+\sqrt{(\mathcal{D}_{\widehat{i}}-\mathcal{D}_{\widehat{i,j}})\mathcal{D}}}\right)+ (i \leftrightarrow j)\,,
			\end{align}
			where coefficient 
			\begin{align}
				N=\frac{(\boldsymbol{C}^{T}\boldsymbol{A}^{-1})_{\widehat{i};j-\theta(j-i)}}{\sqrt{|\boldsymbol{A}_{\widehat{i,j}}|}}\frac{\sqrt{|\boldsymbol{A}|}(\boldsymbol{C}^{T}\boldsymbol{A}^{-1})_i}{\sqrt{(\mathcal{D}_{\widehat{i}}-\mathcal{D})(\mathcal{D}_{\widehat{i}}-\mathcal{D}_{\widehat{i,j}})}}=-\sqrt{-1}\,,
			\end{align}
			after using  \eqref{eq:pref}. When $i\in(k_1,k_2,k_3)$ or/and $j\in(k_1,k_2,k_3)$, we will have the similar expression with the understanding that now $c_0,\boldsymbol{A}, \boldsymbol{C}$ are read out from the $F$, which is obtained from $\mathcal{F}$ (see \eqref{eq:Fexpr}) by integrating out proper variables, for example, when $i=k_1$, integrating out $x_{k_1'}$ instead of $x_{k_1}$.

		\end{itemize}
		
		\subsection{The case of $n=2m+1$}
		
		By combining \eqref{F-c0-1}, \eqref{F-ci-1}
		and \eqref{F-cij-1}, we have 
		\begin{align}
			d{\cal I}_{2m+1} = c_{2m+1\to 2m+1}{\cal I}_{2m+1}+\sum_i c_{2m+1\to 2m;i}  {\cal I}_{2m}^{(i)}+\sum_{i\neq j} c_{2m+1\to 2m-1;ij}  {\cal I}_{2m-1}^{(ij)}\label{In-diff-2}
		\end{align}
		The situation is similar to the case $n=2m$, so we will be brief. For coefficient $c_{2m+1\to 2m+1}$ we have 
		\bea c_{2m+1\to 2m+1} & = & -2\epsilon d\log\mathcal{D}~~~\label{4.23}\eea
		For coefficients to sub-sectors we have 
		\begin{align}\label{eq:2madd122m} 
			c_{2m+1\to 2m;i} & =  \frac{-\epsilon (\boldsymbol{C}^T \boldsymbol{A}^{-1})_i|\boldsymbol{A}|^{1/2}}{2\sqrt{(\mathcal{D}_{\widehat{i}}-\mathcal{D})}|\boldsymbol{A}_{\widehat{i}}|^{1/2}}d\log\left(\frac{\sqrt{-1}(\boldsymbol{C}^T \boldsymbol{A}^{-1})_i|\boldsymbol{A}|^{1/2}/|\boldsymbol{A}_{\widehat{i}}|^{1/2}-\sqrt{\mathcal{D}_{\widehat{i}}}}{\sqrt{-1}(\boldsymbol{C}^T \boldsymbol{A}^{-1})_i|\boldsymbol{A}|^{1/2}/|\boldsymbol{A}_{\widehat{i}}|^{1/2}+\sqrt{\mathcal{D}_{\widehat{i}}}}\right)\notag\\
			&=-\frac{\epsilon \sqrt{-1}}{2}d\log\left(\frac{\sqrt{-1}(\boldsymbol{C}^T \boldsymbol{A}^{-1})_i|\boldsymbol{A}|^{1/2}/|\boldsymbol{A}_{\widehat{i}}|^{1/2}-\sqrt{\mathcal{D}_{\widehat{i}}}}{\sqrt{-1}(\boldsymbol{C}^T \boldsymbol{A}^{-1})_i|\boldsymbol{A}|^{1/2}/|\boldsymbol{A}_{\widehat{i}}|^{1/2}+\sqrt{\mathcal{D}_{\widehat{i}}}}\right)\,,
		\end{align}
		where the case $i\in (k_1,k_2,k_3)$ should be understood similarly. For coefficients to sub-sub-sectors we have 
		\bea  c_{2m+1\to 2m-1;ij}  =  \frac{\sqrt{-1}}{2}\epsilon d \log \left(\frac{\sqrt{-1}(\text{adj}\boldsymbol{A})_{i,j}-\sqrt{|\boldsymbol{A}_{\widehat{i,j}}|}\sqrt{| \boldsymbol{A}| }}{\sqrt{-1}(\text{adj}\boldsymbol{A})_{i,j}+\sqrt{| \boldsymbol{A}_{\widehat{i,j}}| }\sqrt{| \boldsymbol{A}| }}\right)~~~~~~~~\label{4.25}\eea
		where  $\text{adj}\boldsymbol{A}$  denotes adjugate matrix of $\boldsymbol{A}$.

		\section{\label{sec:exam} Degenerate examples}
		
		In the above sections, we have discussed the non-degenerate case systematically. In this section, we 
		will briefly discuss the degenerate cases. There are two degenerate situations:
		\begin{itemize}
			\item[1)]  $|\boldsymbol{A}|=0$.\\
			Multiplying both sides of the equation \eqref{F-c0-1} by $|\boldsymbol{A}|\mathcal{D}$ yields
			\begin{align}
				(\gamma+2)|\boldsymbol{A}|\mathcal{D}F^{\gamma+1}=\boldsymbol{C}^T(\text{adj}\boldsymbol{A})\boldsymbol{\bar{\delta}}F^{\gamma+2}\,.
			\end{align}
			Although $|\boldsymbol{A}|=0$, while $|\boldsymbol{A}|\mathcal{D}=\boldsymbol{C}^T(\text {adj}\boldsymbol{A})\boldsymbol{C}-|\boldsymbol{A}|2c_0=\boldsymbol{C}^T(\text {adj}\boldsymbol{A})\boldsymbol{C}$ remains non-vanishing. Consequently
			\begin{align}\label{eq:detAvanish}
				(\gamma+2)F^{\gamma+1}=\frac{\boldsymbol{C}^T(\text{adj}\boldsymbol{A})\boldsymbol{\bar{\delta}}}{\boldsymbol{C}^T(\text {adj}\boldsymbol{A})\boldsymbol{C}}F^{\gamma+2}\,.
			\end{align}
			\item[2)]  $|\mathcal{D}|=0$.\\
			Multiplying both sides of the equation \eqref{F-c0-1} by $|\mathcal{D}|$ yields
			\begin{align}
				0=-(2\gamma+n-1)F^{\gamma+1}+\boldsymbol{C}^T\boldsymbol{A}^{-1}\boldsymbol{\bar{\delta}}F^{\gamma+1}+\delta_{XYZ}F^{\gamma+1}\,.
			\end{align}
			Thus,
			\begin{align}\label{eq:dvanish}
				F^{\gamma+1}=\frac{\boldsymbol{C}^T\boldsymbol{A}^{-1}\boldsymbol{\bar{\delta}}}{2\gamma+n-1}F^{\gamma+1}+\frac{\delta_{XYZ}}{2\gamma+n-1}F^{\gamma+1}\,.
			\end{align}
		\end{itemize}
		
		In this section, we present several examples to illustrate how to handle degenerate cases.
		For convenience, we use $(a,b,c)$ to denote a diagram in which one branch contains $a$ propagators, while the other two branches contain $b$ and $c$ propagators, respectively. 
		
		\subsection{(3,1,1)}
		
		Without loss of generality, we choose $a_{k_1}=x_3, a_{k_2}=y_4$ and $a_{k_3}=z_5$. In this case,
		\begin{align}
			\gamma&=-2-2\epsilon,~~~~~~\boldsymbol{A}=\begin{pmatrix}
				2c_{1,1} & 2c_{1,2}\\
				2c_{1,2} & 2c_{2,2}
			\end{pmatrix}\notag\\
			\mathcal{D}&=-\frac{ c_{1}^2 c_{2,2}-2c_{2} c_{1} c_{1,2}+c_{2}^2 c_{1,1}+4c_{0} \left(c_{1,2}^2- c_{1,1} c_{2,2}\right)}{2c_{1,2}^2-2 c_{1,1} c_{2,2}}\,.
		\end{align}
		If $|\boldsymbol{A}_{\widehat{1}}|=2c_{2,2}=0$, then using \eqref{eq:detAvanish}, we obtain:
		\begin{align}\label{eq:2112111}
			-2\epsilon\delta(x_1)F^{-1-2\epsilon}&=\frac{(\boldsymbol{C}_{\widehat{1}}^T(\text{adj}\boldsymbol{A}_{\widehat{1}}))_1}{\boldsymbol{C}_{\widehat{1}}^T(\text{adj}\boldsymbol{A}_{\widehat{1}})\boldsymbol{C}_{\widehat{1}}}\delta(x_1)(\delta(x_3)-\delta(x_2))F^{-2\epsilon}\notag\\
			&=\frac{1}{c_2}\delta(x_1)(\delta(x_3)-\delta(x_2))F^{-2\epsilon}\,.
		\end{align}
		Using the master integrals definition \eqref{Ieven} and \eqref{Iodd}, we can get
		\begin{align}\label{eq:dege}
			\mathcal{I}_{4}^{(1)}=\frac{-1}{c_2}(\boldsymbol{C}_{\widehat{1}}^T(\text{adj}\boldsymbol{A}_{\widehat{1}})\boldsymbol{C}_{\widehat{1}})^{1/2}(\mathcal{I}_{3}^{(13)}-\mathcal{I}_{3}^{(12)})=-\mathcal{I}_{3}^{(13)}+\mathcal{I}_{3}^{(12)}\,.
		\end{align}
		We can directly apply the differential equations \eqref{F-c0-1}, \eqref{F-ci-1} and \eqref{F-cij-1} to obtain the CDEs in degenerate cases. Since only $\mathcal{I}_{3}^{(13)}$ and $\mathcal{I}_{3}^{(12)}$ are affected, we include only the $\mathcal{I}_{3}^{(12)}$ term in the differential equations for simplicity (we do not explicitly write both terms, as their treatment follows a similar approach). From \eqref{F-c0-1}, we can get 
		\begin{align}
			\partial_{c_0}(\epsilon^3(2\epsilon)|\boldsymbol{A}|^{1/2}F^{-1-2\epsilon})
			&=-2\epsilon^4\frac{(\boldsymbol{C}^T \boldsymbol{A}^{-1})_1|\boldsymbol{A}|^{1/2} }{\mathcal{D}}\delta(x_1)F^{-2\epsilon}\notag\\
			&=\frac{2\epsilon^3\sqrt{-1}c_{1,2}^2}{c_{2}^2 c_{1,1}-2c_{1} c_{2} c_{1,2}+4c_{0} c_{1,2}^2}\delta(x_1)\delta(x_2)F^{-2\epsilon}\,
		\end{align}
		where to get the second line, we have used \eqref{eq:2112111}. Analogously, we can get
		\begin{align}
			\partial_{c_1}(\epsilon^3(2\epsilon)|\boldsymbol{A}|^{1/2}F^{-1-2\epsilon})
			&=\frac{-\epsilon^3\sqrt{-1}c_{2}c_{1,2}}{c_{2}^2 c_{1,1}-2c_{1} c_{2} c_{1,2}+4c_{0} c_{1,2}^2}\delta(x_1)\delta(x_2)F^{-2\epsilon}\notag\\
			\partial_{c_2}(\epsilon^3(2\epsilon)|\boldsymbol{A}|^{1/2}F^{-1-2\epsilon})
			&=\frac{\epsilon^3\sqrt{-1}c_{1,2} (c_{1} c_{2}-4 c_{0} c_{1,2})}{ c_{2} \left(4c_{0} c_{1,2}^2+c_{2} (c_{1,1} c_{2}-2c_{1} c_{1,2})\right)}\delta(x_1)\delta(x_2)F^{-2\epsilon}\notag\\
			\partial_{c_{1,1}}(\epsilon^3(2\epsilon)|\boldsymbol{A}|^{1/2}F^{-1-2\epsilon}) 
			&=\frac{\epsilon^3\sqrt{-1}c_{1,2} (c_{1} c_{2}-2 c_{0} c_{1,2})}{c_{1,1} \left(4c_{0} c_{1,2}^2+c_{2} (c_{1,1} c_{2}-2c_{1} c_{1,2})\right)}\delta(x_1)\delta(x_2)F^{-2\epsilon}\notag\\
			\partial_{c_{1,2}}(\epsilon^3(2\epsilon)|\boldsymbol{A}|^{1/2}F^{-1-2\epsilon}) 
			&=\frac{\epsilon^3\sqrt{-1}(c_{1} c_{2}-4 c_{0} c_{1,2})}{2 \left(4c_{0} c_{1,2}^2+c_{2} (c_{1,1} c_{2}-2c_{1} c_{1,2})\right)}\delta(x_1)\delta(x_2)F^{-2\epsilon}\,.
		\end{align}
		Recall that $\mathcal{I}_{5}=\epsilon^3(2\epsilon)|\boldsymbol{A}|^{1/2}F^{-1-2\epsilon}$ and $\mathcal{I}_{3}^{(12)}=\epsilon^2\delta(x_1)\delta(x_2)F^{-2\epsilon}$. From above equations, we can easily get the coefficient $c_{5\to 3;12}$ in the degenerate case is
		\begin{align}\label{eq:degecoeff}
			\frac{\sqrt{-1}\epsilon}{2}d\log\left(\frac{c_{2}^2 c_{1,1}-2c_{1} c_{2} c_{1,2}+4c_{0} c_{1,2}^2}{c_{2}^2 c_{1,1}}\right)\,.
		\end{align}

		Alternatively, we can attempt to derive the degenerate CDEs from the non-degenerate ones. From \eqref{eq:dege}, we know that in the degenerate case, $\mathcal{I}_{4}^{(1)}$ decomposes into $\mathcal{I}_{3}^{(12)}$ and $\mathcal{I}_{3}^{(13)}$. Consequently, there are two contributions for the coefficients of $\mathcal{I}_{3}^{(12)}$: 1) A contribution already existed  for the non-degenerate case;
		2) the contribution originally came from the $\mathcal{I}_{4}^{(1)}$, which is further reduced to  $\mathcal{I}_{3}^{(12)}$. Thus, the coefficient for $\mathcal{I}_{5}\to\mathcal{I}_{3}^{(12)}$ is given by
		\begin{align}\label{eq:nondege2dege}
			&c_{5\to 3;12}+c_{5\to 4;1}c_{4;1\to3;12}\notag\\
			&=\frac{\sqrt{-1}\epsilon}{2}d\log\left(\frac{-\sqrt{-4}c_{1,2}-\sqrt{4c_{1,1}c_{2,2}-4c_{1,2}^2}}{-\sqrt{-4}c_{1,2}+\sqrt{4c_{1,1}c_{2,2}-4c_{1,2}^2}}\right)\notag\\ &~~~~~-\frac{\sqrt{-1}\epsilon}{2}d\log\left(\frac{\sqrt{-1}(\boldsymbol{C}^T\boldsymbol{A}^{-1})_1\sqrt{|\boldsymbol{A}|}/\sqrt{|\boldsymbol{A}_{\widehat{1}}|}-\sqrt{\mathcal{D}_{\widehat{1}}}}{\sqrt{-1}(\boldsymbol{C}^T\boldsymbol{A}^{-1})_1\sqrt{|\boldsymbol{A}|}/\sqrt{|\boldsymbol{A}_{\widehat{1}}|}+\sqrt{\mathcal{D}_{\widehat{1}}}}\right)(1)\notag\\
			&=\frac{\sqrt{-1}\epsilon}{2}d\log\left(\frac{\sqrt{-1}c_{1,2}+\sqrt{c_{1,1}c_{2,2}-c_{1,2}^2}}{\sqrt{-1}c_{1,2}-\sqrt{c_{1,1}c_{2,2}-c_{1,2}^2}}\frac{\sqrt{-1}(\boldsymbol{C}^T\boldsymbol{A}^{-1})_1\sqrt{|\boldsymbol{A}|}/\sqrt{|\boldsymbol{A}_{\widehat{1}}|}+\sqrt{\mathcal{D}_{\widehat{1}}}}{\sqrt{-1}(\boldsymbol{C}^T\boldsymbol{A}^{-1})_1\sqrt{|\boldsymbol{A}|}/\sqrt{|\boldsymbol{A}_{\widehat{1}}|}-\sqrt{\mathcal{D}_{\widehat{1}}}}\right)\notag\\
			&=\frac{\sqrt{-1}\epsilon}{2}d\log\left(\frac{c_{2}^2 c_{1,1}-2c_{1} c_{2} c_{1,2}+4c_{0} c_{1,2}^2}{c_{2}^2 c_{1,1}}\right)\,.
		\end{align}
		To obtain the third equation, expand the terms inside the parentheses as a power series in $c_{22}$ and extract the coefficient of the constant term. The result is consistent with \eqref{eq:degecoeff}.
		
		If $\mathcal{D}_{\widehat{i}}=0$, $\mathcal{I}_{2m}^{(i)}=0$ follows from \eqref{Ieven}, and $c_{2m+1\to 2m;i}=0$ follows from \eqref{eq:2madd122m}.  This implies that $\mathcal{I}_{2m}^{(i)}$ can be omitted, while the other terms in the canonical differential equations remain unaffected.
		
		\subsection{(4,1,1)}
		
		In $(4,1,1)$ diagram, $\mathcal{I}_{6}=2\epsilon^4(1+2\epsilon)|\boldsymbol{A}|^{1/2}\mathcal{D}^{1/2}F_{6}^{-2-2\epsilon}$, $\mathcal{I}_{5}^{(1)}=2\epsilon^4|\boldsymbol{A}_{\widehat{1}}|^{1/2}F_{5}^{-1-2\epsilon}$ and $\mathcal{I}_{4}^{(12)}=2\epsilon^3|\boldsymbol{A}_{\widehat{1,2}}|^{1/2}\mathcal{D}_{\widehat{1,2}}F_{4}^{-1-2\epsilon}$. Without loss of generality, we choose $a_{k_1}=x_4, a_{k_2}=y_4$ and $a_{k_3}=z_5$. We derive the degenerate CDEs from the non-degenerate ones. For $\mathcal{D}_{\widehat{1}}=0$, from \eqref{eq:dvanish} we can obtain 
		\begin{align}
			2\epsilon^4\delta(x_1)F^{\gamma+2}=\frac{1}{4}(2\epsilon^3)(\boldsymbol{C}_{\hat{1}}^T \boldsymbol{A}_{\hat{1}}^{-1})_1\delta(x_1)\delta(x_2)F^{\gamma+2}\,+....
		\end{align}
		Consequently,
		\begin{align}
			\mathcal{I}_{5}^{(1)}=\frac{1}{4}\frac{|\boldsymbol{A}_{\widehat{1}}|^{1/2}(\boldsymbol{C}_{\hat{1}}^T \boldsymbol{A}_{\hat{1}}^{-1})_1}{|\boldsymbol{A}_{\widehat{1,2}}|^{1/2}(\mathcal{D}_{\widehat{1,2}})^{1/2}}\mathcal{I}_{4}^{(12)}+...=\frac{\sqrt{-1}}{4}\mathcal{I}_{4}^{(12)}\,+...
		\end{align}
		Here, for simplicity,  we omit other sub-sub-sector master integrals  on the right-hand side of the above two equations, as their treatment follows a similar approach. Analogous to \eqref{eq:nondege2dege}, the coefficient for $\mathcal{I}_{6}\to\mathcal{I}_{4}^{(12)}$ is given by
		\begin{align}
			&c_{6\to4;12}+c_{6\to5;1}c_{5;1\to4;12}\notag\\
			&=\frac{\epsilon\sqrt{-1}}{2}d\log{\frac{\sqrt{(\mathcal{D}_{\widehat{1}}-\mathcal{D})\mathcal{D}_{\widehat{1,2}}}-\sqrt{(\mathcal{D}_{\widehat{1}}-\mathcal{D}_{\widehat{1,2}})\mathcal{D}}}{\sqrt{(\mathcal{D}_{\widehat{1}}-\mathcal{D})\mathcal{D}_{\widehat{1,2}}}+\sqrt{(\mathcal{D}_{\widehat{1}}-\mathcal{D}_{\widehat{1,2}})\mathcal{D}}}}+(1\leftrightarrow 2)\notag\\ &~~~~~~-\frac{\epsilon2\sqrt{-1}}{4} d\log(\frac{\sqrt{\mathcal{D}-\mathcal{D}_{\widehat{1}}}-\sqrt{\mathcal{D}}}{\sqrt{\mathcal{D}-\mathcal{D}_{\widehat{1}}}+\sqrt{\mathcal{D}}})\notag\\
			&=-\frac{\epsilon\sqrt{-1}}{2} d\log\frac{\mathcal{D}_{\widehat{1,2}}}{\mathcal{D}_{\widehat{1,2}}-\mathcal{D}}+\frac{\epsilon\sqrt{-1}}{2}d\log{\frac{\sqrt{(\mathcal{D}_{\widehat{2}}-\mathcal{D})\mathcal{D}_{\widehat{1,2}}}-\sqrt{(\mathcal{D}_{\widehat{2}}-\mathcal{D}_{\widehat{1,2}})\mathcal{D}}}{\sqrt{(\mathcal{D}_{\widehat{2}}-\mathcal{D})\mathcal{D}_{\widehat{1,2}}}+\sqrt{(\mathcal{D}_{\widehat{2}}-\mathcal{D}_{\widehat{1,2}})\mathcal{D}}}}\,,
		\end{align}
		where 
		\begin{align}
			\mathcal{D}&=\boldsymbol{C}^T\boldsymbol{A}^{-1}\boldsymbol{C}-2c_0\,,
		\end{align}
		with
		\begin{align}
			\boldsymbol{A}=\begin{pmatrix}
				2 c_{1,1} & 2c_{1,2} & 2c_{1,3} \\
				2c_{1,2} & 2 c_{2,2} & 2c_{2,3} \\
				2c_{1,3} & 2c_{2,3} & 2 c_{3,3} \\
			\end{pmatrix}
		\end{align}
		
		If $|\boldsymbol{A}_{\widehat{i}}|=0$, $\mathcal{I}_{2m-1}^{(i)}=0$ follows from \eqref{Iodd}, and $c_{2m\to 2m-1;i}=0$ follows from \eqref{eq:n2n-1}.  This implies that $\mathcal{I}_{2m-1}^{(i)}$ can be omitted, while the other terms in the canonical differential equations remain unaffected.
		
		\subsection{(5,1,1)}
		
		Without loss of generality, we choose $a_{k_1}=x_5, a_{k_2}=y_4$ and $a_{k_3}=z_5$. Following the steps outlined in the previous two subsections, the first step in deriving the canonical differential equations (CDEs) for the degenerate case is to determine the coefficient for the transition $\mathcal{I}_{6}^{(i)} \to \mathcal{I}_{5}^{(ij)}$.
		
		For $|\boldsymbol{A}_{\widehat{1}}|=0$, then using \eqref{eq:detAvanish}, we obtain:
		\begin{align}
			\mathcal{I}_{6}^{(1)} = \mathcal{I}_{5}^{(12)}+\ldots\,,
		\end{align}
		where 
		\begin{align}\mathcal{I}_{6}^{(1)} =(1+2\epsilon)2\epsilon^2\delta(x_1)|\boldsymbol{A}_{\widehat{1}}|^{1/2}\mathcal{D}_{\widehat{1}}^{1/2}F^{-2-2\epsilon},~~~\mathcal{I}_{5}^{(12)} =\delta(x_1)\delta(x_2)(2\epsilon^4)|\boldsymbol{A}_{\widehat{1,2}}|^{1/2}F^{-2\epsilon}\,.
		\end{align}
		Analogous to \eqref{eq:nondege2dege}, the coefficient for $\mathcal{I}_{7}\to\mathcal{I}_{5}^{(12)}$ is given by
		\begin{align}
			&c_{7\to5;12}+c_{7\to6;1}c_{6;1\to5;12}\notag\\
			&=\frac{\epsilon\sqrt{-1}}{2} d \log \left(\frac{\sqrt{-1}(\text{adj}\boldsymbol{A})_{1,2}-\sqrt{|\boldsymbol{A}_{\widehat{1,2}}|}\sqrt{| \boldsymbol{A}| }}{\sqrt{-1}(\text{adj}\boldsymbol{A})_{1,2}+\sqrt{| \boldsymbol{A}_{\widehat{1,2}}| }\sqrt{| \boldsymbol{A}| }}\right)\notag\\
			&~~~~-\frac{\epsilon \sqrt{-1}}{2}d\log\left(\frac{\sqrt{-1}(\boldsymbol{C}^T\boldsymbol{A}^{-1})_1\sqrt{|\boldsymbol{A}|}/\sqrt{|\boldsymbol{A}_{\widehat{1}}|}-\sqrt{\mathcal{D}}_{\widehat{1}}}{\sqrt{-1}(\boldsymbol{C}^T\boldsymbol{A}^{-1})_1\sqrt{|\boldsymbol{A}|}/\sqrt{|\boldsymbol{A}_{\widehat{1}}|}+\sqrt{\mathcal{D}}_{\widehat{1}}}\right)\notag\\
			&=\frac{\epsilon\sqrt{-1}}{2} d \log\left(\frac{-|\boldsymbol{A}_{\widehat{2}}|\boldsymbol{C}_{\widehat{1}}^T(\text{adj}\boldsymbol{A}_{\widehat{1}})\boldsymbol{C}_{\widehat{1}}}{((\text{adj}\boldsymbol{A})_{1,2})^2\mathcal{D}}\right) \,,
		\end{align}
		where \begin{align}
			\mathcal{D}&=\boldsymbol{C}^T\boldsymbol{A}^{-1}\boldsymbol{C}-2c_0\,,
		\end{align}
		with
		\begin{align}
			\boldsymbol{A}=\begin{pmatrix}
				2 c_{1,1} & 2c_{1,2} &2 c_{1,3} & 2c_{1,4} \\
				2c_{1,2} & 2 c_{2,2} & 2c_{2,3} & 2c_{2,4} \\
				2c_{1,3} & 2c_{2,3} & 2 c_{3,3} & 2c_{3,4}\\
				2c_{1,4} & 2c_{2,4} &  2c_{3,4} & 2c_{4,4}\\
			\end{pmatrix}
		\end{align}

		\section{\label{sec:conc} Conclusion}
		We systematically analyzed the properties of the one-loop-like integrals under the newly proposed HHM representation of Feynman integrals. This includes providing an alternative iterative reduction scheme equivalent to that in \cite{Huang:2024nij}, based on which we derived the canonical basis and canonical differential equations for this function family. We found that its properties are remarkably similar to those of traditional one-loop integrals. Both cases require only two formulas, distinguished by whether the number of propagators is even or odd, to express all canonical master integrals, as given in \eqref{Ieven} and \eqref{Iodd} of our paper. In the non-degenerate case, their canonical differential equations depend on at most two fewer master integrals, making a complete and systematic discussion feasible. The corresponding matrix elements and symbol structures are provided in \eqref{c2mto2m}, \eqref{eq:n2n-1}, and \eqref{eq:n2n-2}, as well as in \eqref{4.23}, \eqref{eq:2madd122m} and \eqref{4.25}. The symbol structure of the one-loop-like integrals in the HHM representation also closely resembles that of traditional one-loop diagrams (see, for example, \cite{Chen:2022fyw}). Additionally, we presented several examples of degenerate cases for the canonical differential equations.
		
		Since the one-loop-like integral family we studied is the first step that needs to be computed for efficient two-loop calculations under the HHM representation \cite{Huang:2024nij}, providing its canonical differential equations is crucial for directly obtaining its analytic results, performing fast numerical computations, and systematically analyzing its singularity structure. This could further facilitate more efficient two-loop and even higher-loop computations under the HHM representation.
		
		We also observe that the differential equations of one-loop-like integrals can be used to establish IBP relations for the full two-loop integrals, i.e.,  it effectively transforms the problem of multi-variable IBP reduction for arbitrary two-loop integrals into a two-variable reduction problem (where three variables $X,Y, Z$ are reduced to two after integrating out a delta function).
		
		As a result, an interesting direction for future exploration is whether our differential equations for the one-loop-like integrals can aid in studying the reduction properties of complete two-loop or higher-loop integrals and help uncover the iterative structure of IBP relations (see, for example,  \cite{Chen:2022jux}). Furthermore, integrating this representation with modern mathematical tools such as computational algebraic geometry \cite{Gluza:2010ws,Larsen:2015ped,Larsen:2016tdk}, intersection theory \cite{Mizera:2017rqa,Mastrolia:2018uzb} and generating function \cite{Feng:2022hyg} are also worth considering in future research.

		\section*{Acknowledgements}
		
		We would like to thank Li-Hong Huang, Rui-Jun Huang, and  Yan-Qing Ma for early participation in the project and many valuable discussions. 
		This work is supported by Chinese NSF funding under Grant No.11935013, No.11947301, No.12047502 (Peng Huanwu Center), No.12247103, No.U2230402, China Postdoctoral Science Foundation No.2022M720386, and Science Foundation of China University of Petroleum, Beijing (No.2462025YJRC019).
		
		\appendix
		
		\section{\label{sec:appe}Independence of the choice of $\boldsymbol{k}$}
		
		In the paper, we derive the canonical basis and canonical differential equation using the $F$, which is obtained from ${\cal F}$ after integrating out three $a_i$'s. Since different choices of $a_i$'s will give different expressions of $F$, it is not obvious that the canonical basis and canonical differential equation will be independent of these choices. In this part, we will prove this point. 
		
		The procedure of integrated out can be represented by an $(n+1)\times (n-2)$ transformation matrix as
		\begin{align}\label{eq:atrans}
			(1, a_1,\ldots,a_n)^{T}=\boldsymbol{b(k)}(1, \boldsymbol{x}_{\widehat{k_1}},\boldsymbol{y}_{\widehat{k_2}},\boldsymbol{z}_{\widehat{k_3}})^{T}\,,
		\end{align}
		where the $\boldsymbol{x}_{\widehat{k_1}}$ represents the remaining parameters along the $X$-branch, i.e., $\widehat{k_1}$ means the $x_{k_1}$ has been removed. Thus from \eqref{eq:Fexpr} and \eqref{eq:Fkexpr} we get
		\begin{align}
			\boldsymbol{M}=\boldsymbol{b(k)}^T\boldsymbol{\mathcal{M}} \boldsymbol{b(k)}
		\end{align}
		Since matrix $\boldsymbol{b(k)}$ is not a square matrix, we want to  rewrite \eqref{eq:atrans} as
		\begin{align}\label{eq:xtrans}
			(1, a_1,\ldots,a_n)^{T}=\boldsymbol{B(k)}(1, \boldsymbol{x}_{\widetilde{k_1}},\boldsymbol{y}_{\widetilde{k_2}},\boldsymbol{z}_{\widetilde{k_3}})^{T}\,.
		\end{align}
		where the $\boldsymbol{x}_{\widetilde{k_1}}$ represents the same $X$-branch, but the $x_{k_1}$ has been set to zero. The elements of the sqaure $(n+1)\times (n+1)$ matrix $\boldsymbol{B(k)}$ are
		\begin{align}
			\boldsymbol{B(k)}_{i,j}=\left\{
			\begin{aligned}
				\delta_{jk},&\quad i\neq k_1+1,k_2+1,k_3+1\\
				X,&\quad i=k_1+1,j=1\\
				Y,&\quad i=k_2+1,j=1\\
				Z,&\quad i=k_3+1,j=1\\
				-1,&\quad i=k_1+1,j=2,\ldots,n_x+1\\
				-1,&\quad i=k_2+1,j=n_x+2,\ldots,n_x+n_y+1\\
				-1,&\quad i=k_3+1,j=n_x+n_y+2,\ldots,n_x+n_y+n_z+1
			\end{aligned}
			\right.
		\end{align}
		There are some properties about $\boldsymbol{B(k)}$,
		\begin{align}
			\boldsymbol{B(k)}^{-1}=\boldsymbol{B(k)},\quad |\boldsymbol{B(k)}|=-1\,.
		\end{align}
		Similarly, we get
		\begin{align}
			\boldsymbol{\mathcal{M}(k)}&=\boldsymbol{B(k)}^{T}\boldsymbol{\mathcal{M}}\boldsymbol{B(k)}\,,
		\end{align}
		Matrices $\boldsymbol{b(k)}$ and $\boldsymbol{B(k)}$ have following relation: $\boldsymbol{B(k)}$ includes three additional columns, specifically the $k_1+1$-th, $k_2+1$-th, and $k_3+1$-th column.
		Thus $\boldsymbol{M}=\boldsymbol{\mathcal{M}(k)_{\widehat{k+1}}}$, i.e., three corresponding  rows and three columns have been removed from
		$\boldsymbol{\mathcal{M}(k)}$. 
		Utilizing Jacobi's  identity \eqref{Jacobi} we have
		\begin{align}
			|\boldsymbol{M}|=|\boldsymbol{\mathcal{M}(k)_{\widehat{k+1}}}|&= |\boldsymbol{\mathcal{M}(k)}||\boldsymbol{\mathcal{M}(k)}^{-1}_{\boldsymbol{k+1},\boldsymbol{k+1}}|\,,
		\end{align}
		where
		\bea |\boldsymbol{\mathcal{M}(k)}|=|\boldsymbol{B(k)}^{T}\boldsymbol{\mathcal{M}}\boldsymbol{B(k)}|=|\boldsymbol{\mathcal{M}}|\eea
		which is independent of $\boldsymbol{k}$ and 
		\begin{align}
			|\boldsymbol{\mathcal{M}(k)}^{-1}_{\boldsymbol{k+1},\boldsymbol{k+1}}|&=
			\begin{vmatrix} 
				\boldsymbol{\mathcal{M}(k)}^{-1}_{k_1+1,k_1+1}&\boldsymbol{\mathcal{M}(k)}^{-1}_{k_1+1,k_2+1}&\boldsymbol{\mathcal{M}(k)}^{-1}_{k_1+1,k_3+1}\\
				\boldsymbol{\mathcal{M}(k)}^{-1}_{k_1+1,k_2+1}&\boldsymbol{\mathcal{M}(k)}^{-1}_{k_2+1,k_2+1}&\boldsymbol{\mathcal{M}(k)}^{-1}_{k_2+1,k_3+1}\\
				\boldsymbol{\mathcal{M}(k)}^{-1}_{k_3+1,k_1+1}&\boldsymbol{\mathcal{M}(k)}^{-1}_{k_3+1,k_2+1}&\boldsymbol{\mathcal{M}(k)}^{-1}_{k_3+1,k_3+1}
			\end{vmatrix}
		\end{align}
		One can easily discover $\boldsymbol{\mathcal{M}(k)}^{-1}_{\boldsymbol{k+1},\boldsymbol{k+1}}$ is independent of $\boldsymbol{k}$. For example
		\begin{align}
			\boldsymbol{\mathcal{M}(k)}^{-1}_{k_1+1,k_1+1}&=\sum_{js}\boldsymbol{B(k)}_{k_1+1,j}\boldsymbol{\mathcal{M}}^{-1}_{j,s}\boldsymbol{B(k)}^{T}_{s,k_1+1}\notag\\
			&=X^2\boldsymbol{\mathcal{M}}^{-1}_{11}-2X\sum_{j=2}^{n_x+1}\boldsymbol{\mathcal{M}}^{-1}_{1j}+\sum_{j,s=2}^{n_x+1} \boldsymbol{\mathcal{M}}^{-1}_{js}\,
		\end{align}
		which is independent of the explicit choice $x_{k_1}$.  Analogously, $|\boldsymbol{A}|=|\boldsymbol{M}_{\widehat{1}}|$ is also independent of $\boldsymbol{k}$. Since $|\boldsymbol{M}|=|\boldsymbol{A}|(-\mathcal{D})$, consequently, $\mathcal{D}$ is also independent of $\boldsymbol{k}$. These arguments have proved results, such as \eqref{eq:n2n-1}, are independent of  $\boldsymbol{k}$.
		
		Now we prove \eqref{eq:n2n-1prime} can be written into the form \eqref{eq:n2n-1}, which is equal to the proof of 
		$\mathcal{D}'=\mathcal{D}(k_1')_{\widehat{k_1}}$, where $\mathcal{D}(k_1')$ is obtained from the replacement of $x_{k_1'}$ rather than $x_{k_1}$. Recall that $\mathcal{D}'$ is obtained through the simultaneous replacement of $x_{k_1},x_{k_2},x_{k_3},x_{k_1'}$. Since the order of replacements does not affect the result, we can conceptualize $\mathcal{D}'$ as being first obtained from the replacement of $x_{k_1}',x_{k_2},x_{k_3}$, followed by the replacement of $x_{k_1}$. Thus
		\begin{align}
			\mathcal{D}'=-\frac{|\boldsymbol{\mathcal{M}}'(k_1)_{\widehat{k_1+1}}|}{|\boldsymbol{\mathcal{A}}'(k_1)_{\widehat{k_1}}|}\,,
		\end{align}
		where $\boldsymbol{\mathcal{M}}'=\boldsymbol{\mathcal{M}}(k_1',k_2,k_3)$ and $\boldsymbol{\mathcal{A}}'=\boldsymbol{\mathcal{A}}(k_1',k_2,k_3)$. Thus we have $\mathcal{D}'=\mathcal{D}(k_1')_{\widehat{k_1}}$.
		
		\section{\label{sec:oneloop}One-loop check}
		
		In the appendix, we will validate our main results at the one-loop level by examining examples of IBP relations and symbols presented in Sections 3 and 4. This validation will cover one-loop integrals up to the pentagon. For the sake of simplicity, this check will be performed numerically using the following mass and kinematic parameters:
		\begin{align}
			\{m_1^2,m_2^2,m_3^2,m_4^2,m_5^2\}&=\left\{31,37,41,43,47\right\}\nnn \{s_{11},s_{12},s_{13},s_{22},s_{23},s_{33},s_{14},s_{24},s_{34},s_{44}\}&=\left\{\frac{1}{2},\frac{3}{5},\frac{7}{11},\frac{13}{17},\frac{19}{23},\frac{27}{29},\frac{53}{59},\frac{61}{67},\frac{71}{73},\frac{79}{83}\right\}
		\end{align}
		where $s_{ij}=p_i\cdot p_j$. We define the following notations for simplicity:
		\begin{align}
			\mathcal{K}_n&=\det G(p_1,\ldots,p_{n-1}),~~~ \mathcal{B}_n(z_1,\ldots,z_n)=\det G(\ell,p_1,\ldots,p_{n-1})\,,\nnn
			\mathcal{B}'_n(z_1,\ldots,z_n)&=\det G(\ell,p_1,\ldots,p_{n-2};p_{n-1},p_1,\ldots,p_{n-2})\,,\nnn
			\mathcal{B}''_n(z_1,\ldots,z_n)&=\det G(\ell,p_1,\ldots,p_{n-2};\ell,p_1,\ldots,p_{n-3},p_{n-2}+p_{n-1})\,,\nnn
			\mathcal{K}'_n&=\det G(p_1,\ldots,p_{n-2};p_1,\ldots,p_{n-3},p_{n-2}+p_{n-1}),
		\end{align} 
		where $G(a_1,\ldots,a_n;b_1,\ldots,b_n)$ represents the Gram matrix with entries $G_{i,j}=a_i\cdot b_j$, and $G(a_1,\ldots,a_n)\equiv G(a_1,\ldots,a_n;a_1,\ldots,a_n)$
		
		The one-loop FI in Feynman parameterization is given by:
		\begin{align}
			I(\nu_1,\ldots,\nu_n;D)=\frac{\Gamma(\nu-D/2)}{\prod_{j=1}^{n}\Gamma(\nu_j)}\int d^n \boldsymbol{a}\,\delta(1-\sum_{j=1}^n a_j^{\nu_j})\,F^{D/2-\nu}\,,
		\end{align}
		where $a_i$ in $F$ has been substituted with $(1-\sum_{j\neq i} a_j)$. Without loss of generality, we set $a_i=a_1$ hereafter. Consequently, in the one-loop case, equation \eqref{F-c0-1} can be reformulated as\footnote{We remind the reader that our definition of F in the differential equations does not include the prefactor gamma function.}
		\begin{align}\label{dimen}
			I(D+2)=\frac{\mathcal{D}}{D-n+1}I(D)-\frac{(\boldsymbol{C}^T \cdot \boldsymbol{A}^{-1})_i}{D-n+1}I_{\widehat{i+1}}(D)+
			\frac{1+\sum_i(\boldsymbol{C}^T \cdot \boldsymbol{A}^{-1})_i}{D-n+1}I_{\widehat{1}}(D)\,.
		\end{align}
		For notation simplicity, $I(D)\equiv I(1,\ldots,1;D)$. Using the Baikov representation, the dimension shift relation can be derived (see \cite{Abreu:2022mfk}),
		\begin{align}\label{dshift}
			I(\nu_1,\ldots,\nu_n;D+2)=\frac{-2}{\mathcal{K}_n(D-n+1)}\mathcal{B}_n(b_1,\ldots,b_n)I(\nu_1,\ldots,\nu_n;D)\,,
		\end{align}
		where $\mathcal{B}$ is obtained by replacing the arguments $z_i$ of Baikov polynomial $\mathcal{B}_n(z_1,\ldots,z_n)$ with operators $b_i$\footnote{As the operators $b_i$ are commute, the $\mathcal{B}_n(b_1,\ldots,b_n)$ is well-defined.}. These operators lower the value of the exponent $\nu_i$, according to:
		\begin{align}
			b_iI(\nu_1,\ldots,\nu_n;D)=I(\nu_1,\ldots,\nu_i-1,\ldots,\nu_n;D)\,.
		\end{align}
		Take the box diagram as an example, we have:
		\begin{align}\label{bvalue}
			\mathcal{K}_4=&\frac{35143}{46406525}\,,\nnn
			\mathcal{B}_4(b_1,\ldots,b_4)=&-\frac{3853 b_1^2}{521594}+\frac{15451 b_1 b_2}{494615}-\frac{13081 b_1 b_3}{623645}+\frac{97 b_1 b_4}{21505}-\frac{1151309 b_1}{28687670}-\frac{49216817 b_2^2}{1262257480}\nnn
			&+\frac{458714 b_2 b_3}{6860095}-\frac{1731 b_2 b_4}{86020}+\frac{62050463 b_2}{1577821850}-\frac{4985989 b_3^2}{137201900}+\frac{5761 b_3 b_4}{215050}\nnn
			&+\frac{1990892 b_3}{71719175}-\frac{19 b_4^2}{3400}-\frac{86409 b_4}{3118225}-\frac{4047481551}{36605466920}
		\end{align}
		Plugging \eqref{bvalue} and into \eqref{dshift} yields:
		\begin{align}
			I(D+2)=&\frac{20237407755}{69301996 (D-3)}I(D)+\frac{63321995}{597431 (D-3)}I_{\widehat{1}}(D)-\frac{62050463 }{597431 (D-3)}I_{\widehat{2}}(D)\notag\\
			&-\frac{43799624 }{597431 (D-3)}I_{\widehat{3}}(D)+\frac{43722954}{597431 (D-3)}I_{\widehat{4}}(D)+\frac{11655325}{597431 (D-3)}I(-1,1,1,1;D)\notag\\
			&+\frac{246084085 }{2389724 (D-3)}I(1,-1,1,1;D)+\frac{114677747 }{1194862 (D-3)}I(1,1,-1,1;D)\nnn
			&+\frac{35268959 }{2389724 (D-3)}I(1,1,1,-1;D)-\frac{49288690}{597431 (D-3)}I_{\widehat{1,2}}(D)+\frac{33094930 }{597431 (D-3)}I_{\widehat{1,3}}(D)\nnn
			&-\frac{7116890 }{597431 (D-3)}I_{\widehat{1,4}}(D)-\frac{105504220 }{597431 (D-3)}I_{\widehat{2,3}}(D)+\frac{63501735}{1194862 (D-3)}I_{\widehat{2,4}}(D)\nnn
			&-\frac{42268457 }{597431 (D-3)}I_{\widehat{3,4}}(D)
		\end{align}
		Using FIRE6 \cite{Smirnov:2019qkx}, we obtain the following relations:
		\begin{align}
			I(-1,1,1,1;D)=&\frac{448079}{211915}I_{\widehat{1,2}}(D)+\frac{64699 }{211915}I_{\widehat{1,4}}(D)-\frac{300863 }{211915}I_{\widehat{1,3}}(D)-\frac{1151309}{423830}I_{\widehat{1}}(D)\nnn
			I(1,-1,1,1;D)=&\frac{19715476}{49216817}I_{\widehat{1,2}}(D)+\frac{42201688 }{49216817}I_{\widehat{2,3}}(D)-\frac{12700347 }{49216817}I_{\widehat{2,4}}(D)+\frac{124100926 }{246084085}I_{\widehat{2}}(D)\nnn
			I(1,1,-1,1;D)=&-\frac{1438910}{4985989} I_{\widehat{1,3}}(D)+\frac{4587140 }{4985989}I_{\widehat{2,3}}(D)+\frac{1837759 }{4985989}I_{\widehat{3,4}}(D)+\frac{43799624 }{114677747}I_{\widehat{3}}(D)\nnn
			I(1,1,1,-1;D)=&\frac{1940}{4807}I_{\widehat{1,4}}(D)+\frac{11522}{4807}I_{\widehat{3,4}}(D)-\frac{8655 }{4807}I_{\widehat{2,4}}(D)-\frac{345636}{139403}I_{\widehat{4}}(D)\,.
		\end{align}
		Combine all, we  get
		\begin{align}\label{fire}
			I(D+2)=&\frac{20237407755}{69301996 (D-3)}I(D)+\frac{63321995}{1194862 (D-3)}I_{\widehat{1}}(D)-\frac{62050463}{1194862 (D-3)}I_{\widehat{2}}(D)\nnn
			&-\frac{21899812}{597431 (D-3 )}I_{\widehat{3}}(D)+\frac{21861477}{597431 (D-3)}I_{\widehat{4}}(D)
		\end{align}
		Subsequently, we employ our formula \eqref{dimen} to compute the reduction coefficients. The relevant matrices are:
		\begin{align}
			\boldsymbol{C}=\begin{pmatrix}
				\frac{11}{2}&\frac{1281}{170}&\frac{7083807}{1247290}
			\end{pmatrix},~~~~\boldsymbol{A}=\begin{pmatrix}
				1 & \frac{11}{5} & \frac{191}{55} \\
				\frac{11}{5} & \frac{419}{85} & \frac{168907}{21505} \\
				\frac{191}{55} & \frac{168907}{21505} & \frac{7883673}{623645} 
			\end{pmatrix}
		\end{align}
		Thus,
		\begin{align}\label{cadvalue}
			\boldsymbol{C}^T\cdot \boldsymbol{A}^{-1}=\begin{pmatrix}
				\frac{62050463}{1194862}&\frac{21899812}{597431}&-\frac{21861477}{597431}
			\end{pmatrix},~~~~\mathcal{D}=\frac{20237407755}{69301996}
		\end{align}
		Plugging \eqref{cadvalue} back into \eqref{dimen}, the result is consistent with \eqref{fire}. For brevity, the detailed derivation is omitted; however, \eqref{F-ci-1} and \eqref{F-cij-1} are also consistent with the results obtained from FIRE6.
		
		Finally, we present the symbol letters. For an odd number of propagators, taking $n=5$ as an example, the results from \cite{Chen:2022fyw} are
		\begin{align}
			M_{5}&= d\log(\frac{-\mathcal{K}_5}{\mathcal{B}_5(0)})= d\log \frac{237922359363012971055400}{8619263785284712094826057}\,,\nnn
			M_{5,4}&= \scalebox{0.95}{$\frac{1}{2}d\log \frac{\mathcal{B}_5'(0)-\sqrt{\mathcal{B}_4(0)\mathcal{K}_5}}{\mathcal{B}_5'(0)-\sqrt{\mathcal{B}_4(0)\mathcal{K}_5}}= \frac{1}{2}d\log \frac{5 \sqrt{1416156411900275164882983086655}-5159419877130114}{-5 \sqrt{1416156411900275164882983086655}-5159419877130114}$}\,, \nnn
			M_{5,3}&= \scalebox{0.95}{$\frac{i}{4}d\log\frac{\mathcal{B}_5''-\sqrt{-\mathcal{K}_5\mathcal{K}_3}}{\mathcal{B}_5''+\sqrt{-\mathcal{K}_5\mathcal{K}_3}}=\frac{i}{4}d\log\frac{100106342929340 \sqrt{1329566125852131308839}-4913564640308427796116463}{3289249639177747909664313}$}\,,
		\end{align}
		where $\mathcal{B}_5(0)\equiv\mathcal{B}_5(0,0,0,0,0)$. Our results, obtained using equations \eqref{4.23}, \eqref{eq:2madd122m}, and \eqref{4.25}, are:
		\begin{align}
			c_{5\to5}=&-\epsilon\, d\log\frac{8619263785284712094826057}{118961179681506485527700}\nnn
			c_{5\to4;5}=&-\epsilon\, \frac{i}{2}d\log \frac{-5 \sqrt{1416156411900275164882983086655}-5159419877130114}{5 \sqrt{1416156411900275164882983086655}-5159419877130114}\,, \nnn
			c_{5\to3;4,5}=&\scalebox{0.99}{$ \epsilon\, \frac{-i}{2}d\log\frac{-100106342929340 \sqrt{1329566125852131308839}-4913564640308427796116463}{3289249639177747909664313}$}\,.
		\end{align}
		The relation between our MIs ${\cal I}_n$ and $g_n$ in \cite{Chen:2022fyw} is
		\begin{align}
			\begin{pmatrix}
				{\cal I}_5\\
				{\cal I}_{4;5}\\
				{\cal I}_{3;45}\\
				{\cal I}_{2;345}\\
				{\cal I}_{1;2345}
			\end{pmatrix}=T_5\begin{pmatrix}
				g_5\\
				g_{4}\\
				g_{3}\\
				g_{2}\\
				g_{1}
			\end{pmatrix}
		\end{align}
		where $T_5=\text{diag} (4, -4 i, 2, 2i,1)$. The transformation relation between the prefactor matrix $L$ of $d\log$ in our results and $L$ in \cite{Chen:2022fyw} is:
		\begin{align}
			T_5^{-1} L T_5=L'\,,
		\end{align}
		This leads to the following consistency checks:
		\begin{align}
			\epsilon\, M_5=c_{5\to5},~~ \epsilon\, M_{5,4}=-i\,c_{5\to4;5}, ~~\epsilon\, M_{5,3}=\frac{1}{2}c_{5\to3;45},
		\end{align}
		Analogously, for an even number of propagators, the results for $n=4$ in \cite{Chen:2022fyw}
		\begin{align}
			M_{4}&= d\log(\frac{-\mathcal{K}_4}{\mathcal{B}_4(0)})= d\log \frac{138603992}{20237407755}\,,\nnn
			M_{4,3}&= \frac{i}{2}\,d\log \frac{\mathcal{B}'_4(0)-\sqrt{-\mathcal{B}_4(0)\mathcal{K}_3}}{\mathcal{B}'_4(0)-\sqrt{-\mathcal{B}_4(0)\mathcal{K}_3}}=\frac{i}{2}\,d\log\frac{230424 \sqrt{384510747345}-167991663947}{88348834283}\,, \nnn
			M_{4,2}&=\scalebox{0.9}{ $\frac{i}{4}d\log\frac{{\cal K}_4'-\sqrt{-{\cal B}_4(0){\cal B}_2(0)}}{{\cal K}'_4+\sqrt{-{\cal B}_4(0){\cal B}_2(0)}}=\frac{i}{4}d\log\frac{163582377430 \sqrt{87385126686090}+5660768306333952421}{5450315792898285671}$}
		\end{align} 
		Using \eqref{c2mto2m}, \eqref{eq:n2n-1}, and \eqref{eq:n2n-2}, our corresponding results are:
		\begin{align}
			c_{4\to4}=&-\epsilon\, d\log\frac{20237407755}{69301996}\nnn
			c_{4\to3;4}=&-\epsilon\, d\log \frac{-230424 \sqrt{384510747345}-167991663947}{88348834283}\,, \nnn
			c_{4\to2;3,4}=&\epsilon\, \frac{i}{2}d\log\frac{-1511810 \sqrt{1023095381755171041402810}+5660768306333952421}{5450315792898285671}
		\end{align}
		The consistency relations are:
		\begin{align}
			\epsilon\, M_4=c_{4\to4},~~ \epsilon\, M_{4,3}=\frac{i}{2}\,c_{4\to3;4}, ~~\epsilon\, M_{4,2}=-\frac{1}{2}c_{4\to2;34}
		\end{align}
		Our results for both $n=5$ and $n=4$ are consistent with those presented in \cite{Chen:2022fyw}.
		\bibliographystyle{JHEP}
		\bibliography{bib.bib}

	\end{document}